\documentclass[aps,pra,reprint,nofootinbib,superscriptaddress,twocolumn,showpacs,showkeys,longbibliography,amsmath,amssymb]{revtex4-2}
\usepackage{graphicx}
\usepackage{dcolumn}
\usepackage{bm}
\usepackage{braket}
\usepackage{subfigure}
\usepackage[colorlinks,bookmarks=false,citecolor=blue,linkcolor=red,urlcolor=blue]{hyperref}
\usepackage[english]{babel}
\usepackage{changes}

\newcommand{\mycustomsize}{\fontsize{8.5}{10}\selectfont}

\begin{document}
\preprint{APS/123-QED}

\title{Noisy Demkov-Kunike model}
\author{Lin Chen}
\affiliation{Department of Physics, Zhejiang Normal University, Jinhua 321004, People's Republic of China}
\author{Zhaoxin Liang}\email[Corresponding author:~] {zhxliang@zjnu.edu.cn}
\affiliation{Department of Physics, Zhejiang Normal University, Jinhua 321004, People's Republic of China}
\date{\today}

\begin{abstract}
The Demkov-Kunike (DK) model, characterized by a time-dependent Rabi coupling $J~\text{sech}(t/T)$ and on-site detuning $\Delta_0+\Delta_1\tanh(t/T)$, has one of the most general forms of an exactly solvable two-state quantum system, and, therefore, it provides a paradigm for coherent manipulations of a qubit’s quantum state. Despite its extensive applications in the noise-free cases, the exploration of the noisy DK model remains limited. Here, we extend the coherent DK model to take into account of a noisy coupling term $J\rightarrow J_{\text{noisy}}(t)$. We consider colored Markovian noise sources represented by the telegraph noise and Gaussian noise. We present exact solutions for the survival probability $Q^{\text{noisy}}_{\text{DK}}$ of the noisy DK model, namely the probability of the system to remain in its initial state. For the slow telegraph noise, we identify parameter regimes where the survival probability $Q^{\text{noisy}}_{\text{DK}}$ is suppressed rather than enhanced by noise. In contrast, for slow Gaussian noise, the noise always enhances the survival probability $Q^{\text{noisy}}_{\text{DK}}$, due to the absorption of noise quanta across the energy gap. This study not only complements the existing research on the noisy Landau-Zener model, but also provides valuable insights for the control of two-level quantum systems.
\end{abstract}

\maketitle
\section{introduction}
\begin{figure}
      \centering
      \includegraphics[width=0.49\textwidth]{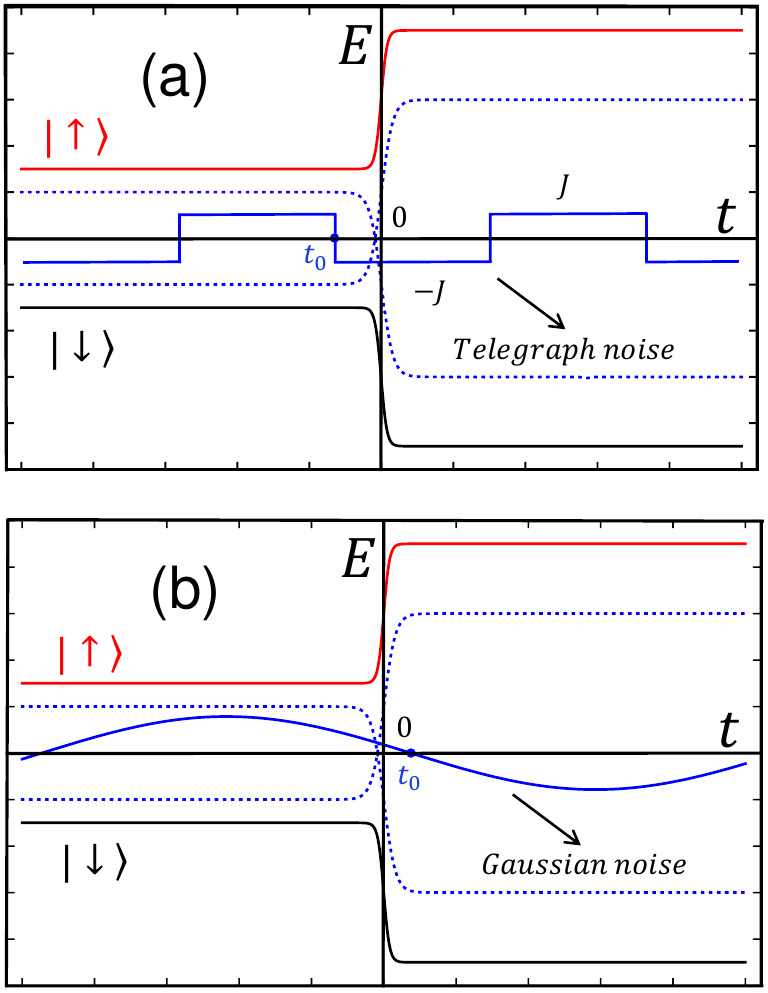}
      \caption{(a) Telegraph noisy Demkov-Kunike (DK) model. In two-state telegraph noise, the stochastic parameter $J_{\text{noisy}}(t)$ in Hamiltonian (\ref{DKH})
switches from $J$ and $-J$ at the time of $t_0$. (b) Gaussian noisy DK model with $J_{\text{noisy}}(t)=J[\tanh(t/T)-\tanh(t_0/T)]$. The time $t=t_0$ where $J_{\text{noisy}}(t)=0$ is close to the time when the level crossing (dashed blue curve) of the diabatic levels closes. }
      \label{Fig1}
\end{figure}
The two-state quantum system not only serves as the building block for quantum information and quantum computation, but also underpins our understanding of various phenomena, such as atomic collisions~\cite{AtomicBook},  molecular magnets~\cite{MagnetBook}, and chemical reactions~\cite{Hanggi1990}. In the study of two-state quantum systems, those models that can be solved analytically are particularly important as they provide benchmarks. Notable examples include
the Landau-Zener (LZ) model~\cite{Landau1932,Zener1932,Wu2000,IVAKHNENKO20231,Kofman2023}, the
Rosen-Zener (RZ) model~\cite{RZ1932}, the Allen-Eberly (AE) model~\cite{AEM1,AEM2},
and the Bambini-Berman (BB)~\cite{BBM} model. In this context, the Demkov-Kunike (DK) model, originally proposed by Ref.~\cite{demkov1969},  represents a rather universal model~\cite{DKM1,DKM2,DKM3}: It can reduce to the RZ, AE, BB, and LZ models under appropriate parameter choices, while avoiding some of their intrinsic drawbacks~\cite{Lacour2007,Simeonov2014}. Indeed, the DK model, in which the Rabi coupling and the on-site detuning depend on time as $J~\text{sech(t/T)}$ and $\Delta_0+\Delta_1\tanh(t/T)$ respectively, provides one of the most general forms of a two-state quantum model that can be analytically solved.

In recent years, the exploration of two-state systems coupled to an environment has attracted considerable attentions. Apart from the fundamental interest in open systems, understanding and controlling noise is also crucial in practical applications~\cite{Sillanpaa2006,Crosson2021}, such as in noisy intermediate-scale quantum~\cite{Preskill2018} and cloud service of quantum
computers~\cite{Cheng2023}. However, the investigation of the impact of noisy environments remains a significant challenge. In general, the noise mainly affects a qubit in two ways~\cite{YanFei2019}, namely, by destroying the superpositions through the randomization of the phase coherence between the two eigenstates, and by generating excitations and altering the state occupations.  Therefore, the aforementioned analytically solvable models need to be revisited to account for the presence of noise. Along this line, the noisy LZ problems have been intensively studied, both theoretically~\cite{Kayanuma1987,Kayanuma1998,Ao1989,Ao1991,Pokrovsky2003,Pokrovsky2007,Nalbach2009,Luo2017,Malla2017} and experimentally~\cite{Harris2008,Quintana2017,Rower2023,dai2022dissipative}. However, to our best knowledge, the study of the noisy DK model still remains elusive. In comparison with the noisy LZ model, the noisy DK model has the distinctive advantage that it may provide a general form of an exactly solvable two-state quantum model coupled to an environment. Hence,
a timely question and fundamental question is: How does the noise influences the quantum transitions in the DK model?

In this work, we investigate the DK tunneling rate in the presence of colored Markovian noises, as exemplified by the telegraph noise and colored Gaussian noise.
Specifically, we focus on the slow noise case along the line of Ref. \cite{Luo2017}, i.e., when the noise correlation time is large compared to typical transition time, where we can derive exact analytical solutions for the survival probability of the system remaining in the initial state. While coupling to classical noise typically results in an enhanced survival probability, we observe the suppression in certain parameter regimes for slow telegraph noise. In contrast, for slow Gaussian noise, we always observed an enhancement of the survival probability, due to the absorption of the noise quanta across the gap. This observation provides valuable insights into the intricate interplay between the noise and the transition dynamics. Our findings not only contribute to the understanding of the noisy DK model but also offer a complementary perspective to the existing studies on the noisy Landau-Zener model. Furthermore, our work introduces new possibilities for the control of two-level quantum systems.

This paper is organized as follows. In Sec.~\ref{NDKM} we describe the noisy DK model. Then, in Secs.~\ref{TDKM} and \ref{GDKM} we systematically investigate how telegraph noise and Gaussian noise affect the tunneling rate of noisy DK model respectively. Finally, in Sec.~\ref{Con} we discuss the experimental conditions for observing the described phenomena, and summarize our work.

\section{Noisy Demkov-Kunike model}\label{NDKM}

The standard DK model, as described in Refs.~\cite{DKM1,DKM2,DKM3}, is characterized by the quasi-linear level-crossing of a bell-shaped pulse with finite detuning.
When noise is considered (see Fig. \ref{Fig1}),  the Hamiltonian takes the form
\begin{equation}
H=\left[\Delta_0+\Delta_1 \tanh\left(t/T\right)\right]\sigma_z+J_{\text{noisy}}(t) \text{sech}\left(t/T\right)\sigma_x,\label{DKH}
\end{equation}
where $\sigma_i$ $(i=x,z)$ are the Pauli matrices. Here, the $\Delta_0$ and $\Delta_1$ are referred to as
the static and chirp detuning parameters, respectively, and $T$ is the scanning period of the external pulse field.  The second term in the Hamiltonian (\ref{DKH})
represents the intrinsic interactions between the two diabatic states ($|\uparrow\rangle$ and $|\downarrow\rangle$), which induces the transitions. The external noise appears as the stochastic parameter~\cite{Ying2015,Zoller1981} $J_{\text{noisy}}(t)$, which fluctuates over time according to the colored Markovian noise sources~\cite{Ying2015,Zoller1981}, as exemplified by the telegraph noise and Gaussian noise. For both types of noise, we have the mean value $\langle J_{\text{noisy}}(t)\rangle=0$ and the first order correlation $\langle J_{\text{noisy}}(t+\tau),J_{\text{noisy}}(t)\rangle=\sigma^2\exp(-|\tau|/\tau_c)$, where $\tau_c$ is the correlation time, $\sigma^2$ is the variance, $\langle \rangle$ denotes the stochastic average and $\langle X, Y\rangle\equiv \langle XY\rangle-\langle X\rangle\langle Y\rangle$. The noisy DK model contains three important noisy models as the special cases, namely
the noisy RZ model~\cite{RZ1932} for $\Delta_1=0$, the noisy AE model~\cite{AEM1,AEM2} for $\Delta_0=0$, and the noisy BB model~\cite{BBM} for $\Delta_0=\Delta_1$.

As a reference, let
us first briefly recapitulate the properties of the noise-free
DK model with $J_{\text{noisy}}(t)=J$. There,
the time evolution of the wave function $\psi(t)=[C_1(t),C_2(t)]^T$ is governed by $i\partial\psi/\partial t=H\psi$. By introducing the new variable $z=(1+\tanh(t/T))/2$, with $z\in[0,1]$ corresponding to $t\in(-\infty,+\infty)$, the solutions of $C_1(t)$ can be transformed into the solutions of the
Gauss hypergeometric equation
\begin{equation}
z(1-z)\frac{d^2C_1}{dz^2}+\left[\nu-\left(\lambda+\mu+1\right)z\right]\frac{dC_1}{dz}-\lambda\mu C_1=0,\label{DKE}
\end{equation}
where $\nu=1/2-iT(\Delta_0-\Delta_1)$, $\lambda=iT[\sqrt{\Delta^2_1-J^2}+\Delta_1]$, and $\mu=iT[\Delta_1-\sqrt{\Delta^2_1-J^2}]$.
Equation (\ref{DKE}) has two linearly independent solutions expressed by the hypergeometric function, i.e., ${}_2F_1\left(\lambda,\mu,\nu,z\right)$ and $z^{1-\nu}{}_2F_1\left(\lambda+1-\nu,\mu+1-\nu,2-\nu,z\right)$, respectively.

The key quantity of interest is the survival probability $Q_{\text{DK}}=|C_1(t\rightarrow \infty)|^2$ under the initial condition $|C_1(t\rightarrow -\infty)|=1$.
For the noise-free DK model, the exact expression of $Q_{\text{DK}}$ was obtained in Refs.~\cite{DKM1,DKM2,DKM3} (or the detailed derivation can be found in Appendix A of Ref.~\cite{DKM3})
\begin{equation}
Q_{\text{DK}}=\frac{\cosh(2\pi T\Delta_1)+\cosh(2\pi T\sqrt{\Delta^2_1-J^2})}{\cosh(2\pi T\Delta_0)+\cosh(2\pi T\Delta_1)}.
\end{equation}
Setting $\Delta_0=0$ yields the transition probability for the AE model, $Q_{\text{AE}}=1-\sinh^2(\pi T\sqrt{\Delta^2_1-J^2})/\cosh^2(\pi T\Delta_1)$~\cite{AEM1,AEM2}. Whereas, if $\Delta_1=0$, one obtains $Q_{\text{RZ}}=\cos^2(\pi T J)/\cosh^2(\pi T\Delta_0)$ for the RZ model~\cite{RZ1932}.

The presence of the noisy component $J_{\text{noisy}}(t) $ in the Hamiltonian parameter
leads to random shaking of system. We denote the survival probability in the presence of noise by $Q_{\text{DK}}^{\text{noisy}}$. Depending on the ratio between the noise correlation time $\tau_c$ and the typical transition time $\tau_{\text{DK}}\propto 1/J$ of the DK model, there are two limits:
(i) In the limit of fast noise $\tau_c/ \tau_{\text{DK}}\rightarrow 0$, the stochastic parameter $J_{\text{noisy}}(t)$ is expected to undergo many oscillations within the transition time, so that the occupations in each of the two diabatic levels are nearly identical. (ii)  In the limit of slow noise $\tau_c/ \tau_{\text{DK}}\rightarrow \infty$, the $J_{\text{noisy}}(t)$ can be treated as a constant on the time scale of transition. This implies that the resulting $Q_{\text{DK}}^{\text{noisy}}$ can be roughly averaged over the distribution $P(J)$ of the stochastic parameter $J_{\text{noisy}}(t)$ as 
$
\langle Q_{\text{DK}}^{\text{noisy}}\rangle=\int dJ P(J)Q_{\text{DK}}^{\text{noisy}}
$.
In our work, we focus on the slow noise case in the sense of (ii), but taking into account of the finite $1/\tau_c$ correction associated with the level crossing regime, along the line of Ref.~\cite{Luo2017}.

We will explore two typical kinds of colored Markovian noises~\cite{Ying2015,Zoller1981}. In Sec.~\ref{TDKM}, we consider a random telegraph process as illustrated in Fig.~\ref{Fig1} (a), and analytically study the survival probability in the DK model in the fast noise limit. In Sec.~\ref{GDKM}
we consider the Gaussian noise where $J_{\text{noisy}}(t)$ changes continuously [c.f. Fig.~\ref{Fig1} (b)] and investigate the transition behavior in the slow noise limit.

\onecolumngrid

\begin{center}
 \begin{figure}[h]
      \centering
      \includegraphics[width=0.96\textwidth]{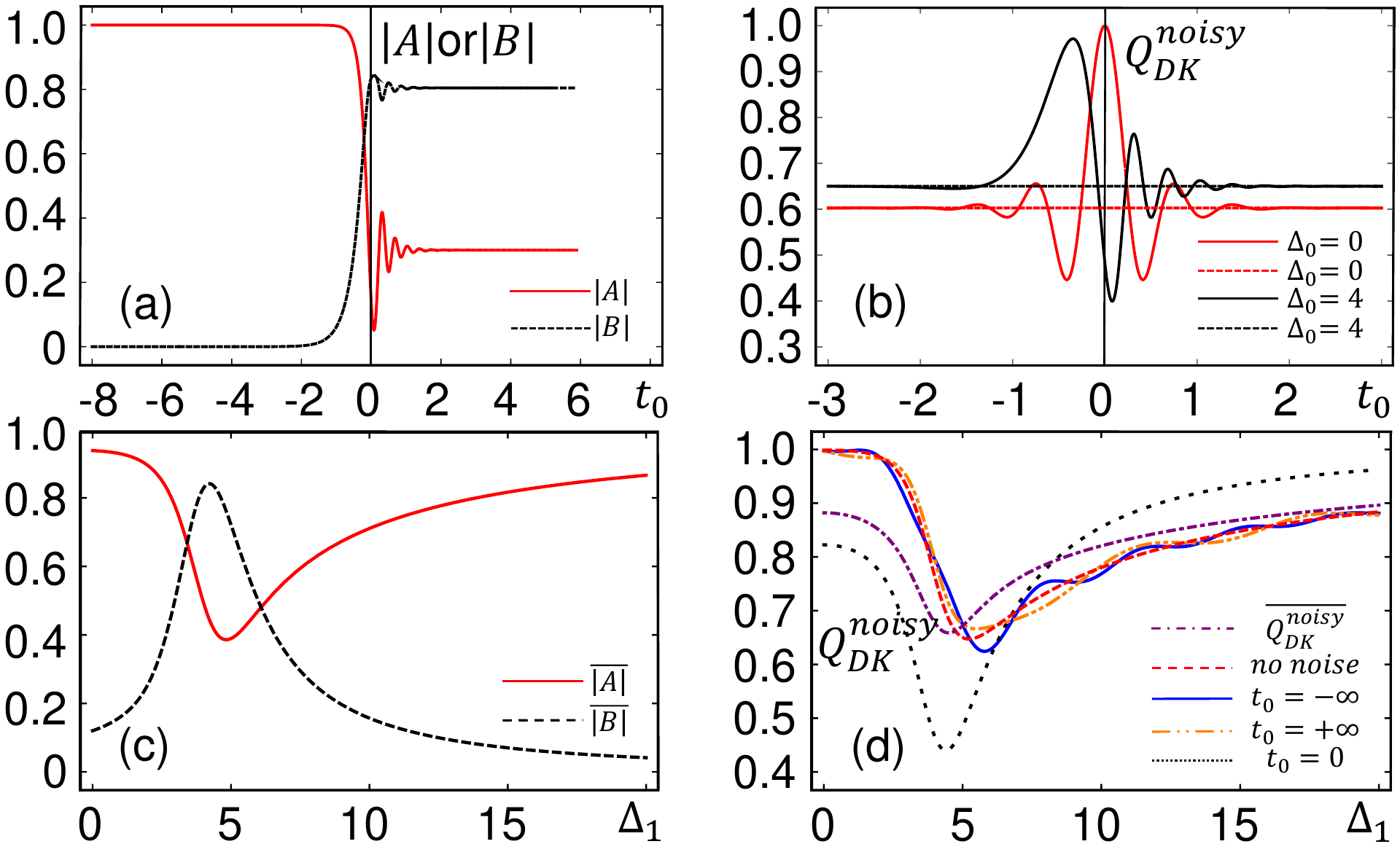}
      \caption{Effects of fast telegraph noise on the tunneling rate $Q^{\text{noisy}}_{\text{DK}}$ of DK model. (a) and (c): The magnitude of the coefficients $A$ in Eq.~(\ref{PA}) and $B$ in Eq.~(\ref{PB}) as a function of (a) the switching time $t_0$ and (c) chirp detuning parameter $\Delta_1$. (b) and (d): Tunneling rate $Q^{\text{noisy}}_{\text{DK}}$ in Eq.~(\ref{DKT}) as a function of (b) $t_0$ and (d) $\Delta_1$. In (b), the dashed curves represent the tunneling rate in the absence of noise. For parameters in each plot, we choose  (a) $J=\pi/2$, $\Delta_0=4$ and $\Delta_1=5$;  (b) $J=\pi/2$ and $\Delta_1=5$; (c) $J=\pi/2$, $\Delta_0=4$ and $t_0=0$; (d) $J=\pi/2$ and $\Delta_0=4$.}
      \label{Fig2}
 \end{figure}
\end{center}

\twocolumngrid
\section{Telegraph noisy DK model}\label{TDKM}


In the two-state telegraph noise model, the stochastic parameter $J_{\text{noisy}}(t)$ in Hamiltonian (\ref{DKH}) randomly switches between two discrete values, $-J$ and $J$. The telegraph noise property is characterized by $\langle J_{\text{noisy}}(t)\rangle=0$ and $\langle J_{\text{noisy}}(t+\tau),J_{\text{noisy}}(t)\rangle=J^2\exp(-|\tau|/\tau_c)$. We follow Ref.~\cite{Luo2017} and consider a sufficiently slow noise but with finite $1/\tau_c$ correction.  During the course of transition, the noise jump typically occurs at time $t_0\sim \tau_c\gg \tau_\textrm{DK}\sim 1/J$ in agreement with the slow noise assumption. However, as illustrated in Fig.~\ref{Fig1} (a), there is some (small) chance that the random switch occurs near the level crossing point that may significantly affects the tunneling probability. 

Below, we exactly solve the dynamics governed by Hamiltonian (\ref{DKH}) for the survival probability $Q^{\text{noisy}}_{\text{DK}}$, under the initial conditions $C_{1}(-\infty)=1$ and $C_{2}(-\infty)=0$. We proceed in two steps. First, we consider $J_{\text{noisy}}(t)$ flip its sign once at some random time $t_0$ during the transition, and calculate the corresponding $Q^{\text{noisy}}_{\text{DK}}$. Then,  since $t_0$ is random, we average $Q^{\text{noisy}}_{\text{DK}}$ over $t_0$ to get the average $ \overline{{Q^{\text{noisy}}_{\text{DK}}}}$.

We begin with calculating the transition dynamics when there occurs one switch between the two discrete values $-J$ and $J$ at some $t_0$.  For times $t<t_0$, the system dynamics are governed by Eq.~(\ref{DKE}) with $J_{\text{noisy}}(t)=J$. Therefore, the instantaneous state can be expressed in terms of Gauss hypergeometric functions as
\begin{equation}
\left(\begin{array}{c}
C_{1}(z)\\
C_{2}(z)
\end{array}\right)=\left(\begin{array}{c}
_{2}F_{1}\left(\lambda,\mu,\nu,z\right)\\
\frac{\sqrt{\lambda \mu z(1-z)}}{\nu}{}_{2}F_{1}\left(\lambda+1,\mu+1,\nu+1,z\right)
\end{array}\right), \label{DKS1}
\end{equation}
where the parameters $\mu$, $\nu$ and $\lambda$ are defined in Eq.~(\ref{DKE}).

For times $t>t_0$, the stochastic parameter $J_{\text{noisy}}(t)$ is
switched to $J_{\text{noisy}}(t)=-J$. In this case, the general solution of Eq.~(\ref{DKE}) involves a linear superposition of two hypergeometric functions, i.e.,
\begin{widetext}
\mycustomsize
\begin{eqnarray}
\left(\begin{array}{c}
C_{1}(z)\\
C_{2}(z)
\end{array}\right)
=A\left(\begin{array}{c}
_{2}F_{1}\left(\lambda,\mu,\nu,z\right)\\
-\frac{\sqrt{\lambda \mu z(1-z)}}{\nu}{}_{2}F_{1}\left(\lambda+1,\mu+1,\nu+1,z\right)
\end{array}\right)
+B\left(\begin{array}{c}
z^{1-\nu}{}_{2}F_{1}\left(\lambda+1-\nu,\mu+1-\nu,2-\nu,z\right)\\
-\sqrt{ \frac{z(1-z)}{\lambda \mu}}(1-\nu)z^{-\nu}{}_{2}F_{1}\left(\lambda+1-\mu,\mu+1-\nu,1-\nu,z\right)
\end{array}\right).\label{DKS2}
\end{eqnarray}

\normalsize

Here, the coefficients $A$ and $B$ are determined by the continuity condition of $C_{1}$ and $C_{2}$ in Eqs.~(\ref{DKS1}) and (\ref{DKS2}) at $t_0$ as
\begin{eqnarray}
A(t_0)&=&\frac{\frac{{}_{2}F_{1}\left(\lambda+1-\nu,\mu+1-\nu,1-\nu,z_0\right)}{{}_{2}F_{1}\left(\lambda+1,\mu+1,\nu+1,z_0\right)}+\frac{\lambda\mu z_0}{\nu(1-\nu)}\frac{{}_{2}F_{1}\left(\lambda+1-\nu,\mu+1-\nu,2-\nu,z_0\right)}{{}_{2}F_{1}\left(\lambda,\mu,\nu,z_0\right)}}{\frac{{}_{2}F_{1}\left(\lambda+1-\nu,\mu+1-\nu,1-\nu,z_0\right)}{{}_{2}F_{1}\left(\lambda+1,\mu+1,\nu+1,z_0\right)}-\frac{\lambda\mu z_0}{\nu(1-\nu)}\frac{{}_{2}F_{1}\left(\lambda+1-\nu,\mu+1-\nu,2-\nu,z_0\right)}{{}_{2}F_{1}\left(\lambda,\mu,\nu,z_0\right)}},\label{PA}\\
B(t_0)&=&\frac{2}{-\frac{\nu(1-\nu)}{\lambda \mu}z^{-\nu}_0\frac{{}_{2}F_{1}\left(\lambda+1-\nu,\mu+1-\nu,1-\nu,z_0\right)}{{}_{2}F_{1}\left(\lambda+1,\mu+1,\nu+1,z_0\right)}+z_0^{1-\nu}\frac{{}_{2}F_{1}\left(\lambda+1-\nu,\mu+1-\nu,2-\nu,z_0\right)}{{}_{2}F_{1}\left(\lambda,\mu,\nu,z_0\right)}}.\label{PB}
\end{eqnarray}
For $t_0\rightarrow \infty$, corresponding to the absence of parameter switching, we obtain $A=1$ and $B=0$ as expected. The introduction of parameter switching leads to $B\neq 0$, causing a significant impact on the survival probability, as demonstrated below.

Using Eq.~(\ref{DKS2}), we obtain an exact expression for the probability to remain at
the same adiabatic level as
\begin{eqnarray}
Q^{\text{noisy}}_{\text{DK}}\!\!&=&|A(t_0)|^2\frac{\cosh(2\pi T\Delta_0)+\cosh(2\pi T\sqrt{\Delta^2_1-J^2})}{\cosh(2\pi T\Delta_0)+\cosh(2\pi T\Delta_1)}\nonumber\\
&+&|B(t_0)|^2\frac{[1-(\Delta_0-\Delta_1)^2T^2][\cosh(2\pi T\Delta_1)-\cosh(2\pi T\sqrt{\Delta^2_1-J^2})]}{J^2T^2[\cosh(2\pi T\Delta_0)+\cosh(2\pi T\Delta_1)]}\nonumber\\
\!\!\!&+&A^*B\frac{\Gamma\left[\frac{1}{2}-\frac{i}{2}(\Delta_0+\Delta_1)T\right]\Gamma\left[\frac{1}{2}-\frac{i}{2}(\Delta_0-\Delta_1)T\right]\Gamma\left[\frac{3}{2}-\frac{i}{2}(\Delta_0-\Delta_1)T\right]\Gamma\left[\frac{1}{2}+\frac{i}{2}(\Delta_0+\Delta_1)T\right]}
{\Gamma\left[\frac{1-T\sqrt{J^2+\Delta^2_1}+i\Delta_0T}{2}\right]\Gamma\left[\frac{1+T\sqrt{J^2+\Delta^2_1}+i\Delta_0T}{2}\right]\Gamma\left[\frac{2-T\sqrt{J^2-\Delta^2_1}-\Delta_1T)}{2}\right]\Gamma\left[\frac{2+T\sqrt{J^2-\Delta^2_1}-\Delta_1T)}{2}\right]}+h.c.,\label{DKT}
\end{eqnarray}
\end{widetext}
where $\Gamma$ is the Gamma function.

Equation (\ref{DKT}) constitutes the first key result of this study, which describes the impact of one random switch on the tunneling rate in the DK model. By setting $A=1$ and $B=0$, Equation (\ref{DKT})
precisely reproduces the noise-free results reported in Refs.~\cite{DKM1,DKM2,DKM3}.

Since in Eq.~(\ref{DKT}) the noise is fully encoded in the coefficients $A(t_0)$ and $B(t_0)$, we now analyze how the telegraph switching time $t_0$ affects $A(t_0)$ and $B(t_0)$, as shown in Fig.~\ref{Fig2} (a). In the asymptotical limit $t_0\rightarrow -\infty$, corresponding to the case with $J_{\text{noisy}}=-J$, it's clear that $A\rightarrow 1$ and $B\rightarrow 0$. This can be understood as follows: (i) the energy gap of DK model is $\sim 2\sqrt{(\Delta_0-\Delta_1)^2+J^2}$ at $t_0\rightarrow -\infty$, (ii) the system is initially prepared in a state with $A=1$ and $B=0$, and $J_{\text{noisy}}$ is switched from $J$ to $-J$, so that the telegraph noise is not strong enough to excite the $B$ state. In contrast, when $t_0\rightarrow 0$, there exists level crossing in the parameter regime $\Delta_0<\Delta_1$ [see dashed curves in Fig.~\ref{Fig1} (a)]. In this case, an arbitrarily small $J$ can excite the system, changing $B$ from zero to nonzero [see black curves in Fig.~\ref{Fig2} (a)]. In Fig.~\ref{Fig2} (c), we further study how the gap closing affects the coefficients $A(0)$ and $B(0)$  at $t_0=0$ by varying $\Delta_1$, when $\Delta_0=4$. We see that the optimal enhancement of $B$ occurs at $\Delta_1\approx 4$, corresponding to where the energy gap almost closes.

Next, based on the behaviors of $A(t_0)$ and $B(t_0)$ under various $t_0$ and $\Delta_1$, we study how the onset of one random jump affects the tunneling rate. In Fig.~\ref{Fig2} (b), we fix $\Delta_1$ and show $Q_{\text{DK}}$ as a function of $t_0$. There, when $\Delta_0=0$ (red curves), the results are similar as that of the telegraph noisy LZ model studied in Ref.~\cite{Luo2017}. Moreover, $Q^{\text{noisy}}_{\text{DK}}$ is symmetric with respect to $t_0$ and exactly recovers the noise-free counterpart in the limit $t_0\rightarrow \pm \infty$. When $\Delta_0\neq 0$ (black curves), however, the $Q^{\text{noisy}}_{\text{DK}}$ becomes asymmetric with respect to $t_0$. In addition, it exactly recovers the noise-free counterpart in the limit of $t_0\rightarrow \pm \infty$. The symmetry can be understood as arising from the symmetry of the energy levels with respect to $t=0$. Surprisingly, we see that $Q^{\text{noisy}}_{\text{DK}}$ decreases with $t_0$ in the regime where $\Delta_0+\Delta_1\tanh(t_0/T)\rightarrow 0$ in Hamiltonian (\ref{DKH}).  

To further understand the noise-suppressed tunneling in the regime $\Delta_0+\Delta_1\tanh(t_0/T)\rightarrow 0$, we fix $t_0=0$ and analyze how the tunneling rate depends on $\Delta_1$. In Fig.~\ref{Fig2} (d),
we show $Q^{\text{noisy}}_{\text{DK}}$ as a function of $\Delta_1$ for different $t_0$, and compare it with the noise-free case (dashed red curves). As expected, both asymptotic results in the limit $t_0\rightarrow  -\infty$ (solid blue curves) and the limit $t_0\rightarrow  +\infty$ (dotted-dashed orange curves) almost coincide with the noise-free case. In contrast, the result of the noisy DK model for $t_0\rightarrow 0$ (dotted black curves) differs significantly from the noise-free case. In particular, there is a dip of $Q^{\text{noisy}}_{\text{DK}}$ where $\Delta_1$ corresponds to the level crossing closing. Thus we conclude that the smaller is the energy gap of the telegraph noisy DK model, the stronger suppression of $Q^{\text{noisy}}_{\text{DK}}$ is. 

Finally, we account for the random nature of $t_0$ and average Eq.~(\ref{DKT}) over $t_0$~\cite{Luo2017} to obtain the corresponding results. The average result $\overline{Q^{\text{noisy}}_{\text{DK}}}$ is plotted as the purple dashed-dotted curve in Fig. \ref{Fig2} (d). Notice that $\overline{Q^{\text{noisy}}_{\text{DK}}}$ shows qualitatively similarity as the result for a single $t_0\sim 0$. This suggests that the transition can be particularly strongly affected by a random occurrence of switch near the level crossing. 

\section{Gaussian noisy DK model}\label{GDKM}

So far, we have systematically studied how the fast telegraph noise affects the DK tunneling rate $Q^{\text{noisy}}_{\text{DK}}$ based on Eq.~(\ref{DKT}). In this section, we consider slow Gaussian noise characterized by $\langle J_{\text{noisy}}(t)\rangle=0$ and $\langle J_{\text{noisy}}(t+\tau),J_{\text{noisy}}(t)\rangle=J^2\exp(-|\tau|/\tau_c)$, and investigate
its effect on $Q^{\text{noisy}}_{\text{DK}}$. Note that Ref. \cite{Kayanuma1985} has first investigated effect of the slow Gaussian noise on the tunneling rate in the context of LZ model. Here, extending the approach developed in Refs. \cite{Luo2017,Kayanuma1985} for the study of the LZ model with slow Gaussian noise, we seek to exactly solve the Gaussian noisy DK model.

Specifically, we shall assume the following form for the off-diagonal term of Hamiltonian (\ref{DKH})
\begin{equation}
J_{\text{noisy}}(t)\text{sech}(t/T)=J\left[\tanh\left(\frac{t}{T}\right)-\tanh\left(\frac{t_0}{T}\right)\right],\label{Gnoise}
\end{equation}
with $t_0$ being a random number.
Similar as the case of telegraph noise, the relevant situation is expected to be when $t_0$ is near the level crossing closing.

\begin{figure}
      \centering
     \includegraphics[width=0.48\textwidth]{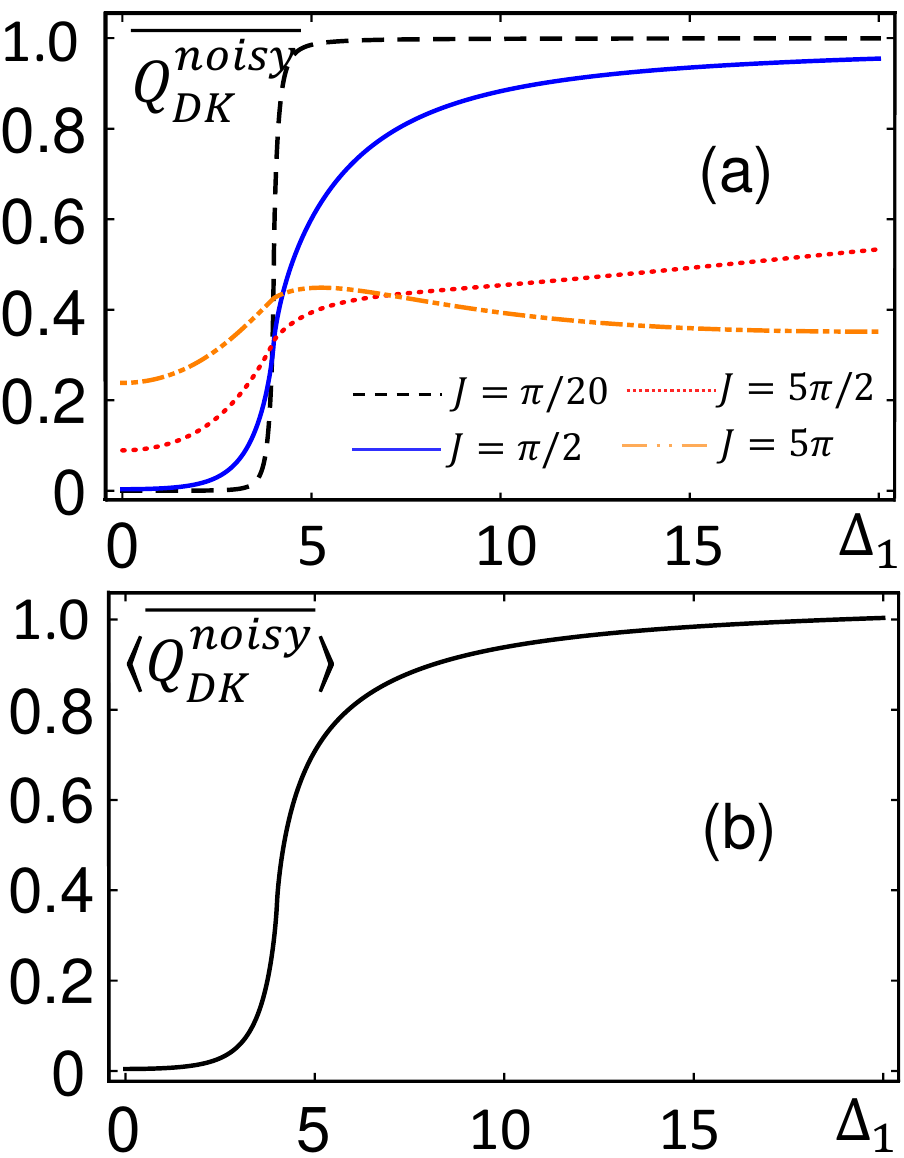}
      \caption{Super-adiabaticity of Gaussian noisy DK model. (a) Time-averaged tunneling rate ${\overline{Q^{\text{noisy}}_{\text{DK}}}}=\int^{+\infty}_{-\infty} dt_0/\tau_c Q^{\text{noisy}}_{\text{DK}}$ as a function of $\Delta_1$. (b) Noise-averaged tunneling rate $\langle{\overline{Q^{\text{noisy}}_{\text{DK}}}}\rangle=\int dJ P(J){\overline{Q^{\text{noisy}}_{\text{DK}}}}$ as a function of $\Delta_1$. Here, $P(J)$ is the Gaussian-type distribution. In both plots, we use $\Delta_0=4$ and $\tau_c=1$.}
      \label{Fig3}
\end{figure}

We start with considering a single choice of $t_0$. Using Eq.~(\ref{Gnoise}), the dynamics of the Gaussian noisy DK model is governed by two-coupled equations
\begin{widetext}
\begin{eqnarray}
i\frac{d C_{1}}{dt}	&=&\left[\Delta_{0}+\Delta_{1}\tanh\left(\frac{t}{T}\right)\right]C_{1}+J\left[\tanh\left(\frac{t}{T}\right)-\tanh\left(\frac{t_0}{T}\right)\right]C_{2},\label{DKGE1}\\
i\frac{d C_{2}}{dt}	&=&J\left[\tanh\left(\frac{t}{T}\right)-\tanh\left(\frac{t_0}{T}\right)\right]C_{1}-\left[\Delta_{0}+\Delta_{1}\tanh\left(\frac{t}{T}\right)\right]C_{2}.\label{DKGE2}\label{GDKE}
\end{eqnarray}
Equations (\ref{GDKE}) can be exactly solved by introducing the following new variables,
\begin{equation}
\left(\begin{array}{c}
C_{1}^{\prime}\\
C_{2}^{\prime}
\end{array}\right)=\left(\begin{array}{cc}
\cos\theta & \sin\theta\\
\sin\theta & -\cos\theta
\end{array}\right)\left(\begin{array}{c}
C_{1}\\
C_{2}
\end{array}\right)\label{Tvariable}
\end{equation}
with $\tan (2\theta)=J/\Delta_1$. Utility of Eq.~(\ref{Tvariable}) transforms Eq.~(\ref{GDKE}) into the dynamical equations for $C_1^\prime$ and $C_2^\prime$, i.e.,
\begin{eqnarray}
i\frac{d C^\prime_{1}}{dt}	&=&\left[\Delta^\prime_0+\Delta^\prime_1\tanh\left(\frac{t}{T}\right)\right]C^\prime_{1}+J^\prime C^\prime_{2},\label{DKGE1}\\
i\frac{d C^\prime_{2}}{dt}	&=&J^\prime C^\prime_{1}-\left[\Delta^\prime_0+\Delta^\prime_1\tanh\left(\frac{t}{T}\right)\right]C^\prime_{2},\label{DKGE2}
\end{eqnarray}
with $\Delta^\prime_0=(\Delta_{0}\cos^22\theta-\Delta_1\sin^22\theta\tanh(t _0/T))/\cos2\theta $, $\Delta^\prime_1=\Delta_{1}/\cos2\theta$ and $J^\prime=[\Delta_0+\Delta_1\tanh\left({t_0}/{T}\right)]\sin 2\theta$.

It turns out that Eqs.~(\ref{DKGE1}) and (\ref{DKGE2}) are the Demkov-Kunike-II model~\cite{Luo2017} with the
renormalized parameters $\Delta^\prime_0$, $\Delta^\prime_1$ and $J^\prime$. After straightforward yet tedious calculations, therefore, we analytically obtain the exact result
\begin{equation}
Q^{\text{noisy}}_{\text{DK}}(t_0,J)=\frac{\sinh[\pi T(E_e-E_a+2\Delta^\prime_1)/2]\sinh[\pi T(E_a-E_e+2\Delta^\prime_1)/2]}{\sinh(\pi T E_a)\sinh(\pi T E_e)},\label{GDKT}
\end{equation}
with $E_a=\sqrt{(\Delta_0^\prime-\Delta^\prime_1)^2+J^{\prime 2}}$ and $E_e=\sqrt{(\Delta_0^\prime+\Delta_1^\prime)^2+J^{\prime 2}}$.
\end{widetext}
Equation (\ref{GDKT}) is another key result of this study, which describes the survival probability of the system at the initial level for the Gaussian noisy DK model.

Since $t_0$ is random, next we average $Q^{\text{noisy}}_{\text{DK}}(t_0,J)$ in Eq.~(\ref{GDKT}) over
$t_0$ as ${\overline{Q^{\text{noisy}}_{\text{DK}}}}=\int^{+\infty}_{-\infty} dt_0/\tau_c Q^{\text{noisy}}_{\text{DK}}$. Consequently, relevant for the tunneling rate are three free parameters $\Delta_0$, $\Delta_1$ and $J$.  In Fig.~\ref{Fig3} (a), we show ${\overline{Q^{\text{noisy}}_{\text{DK}}}}$ as a function of $\Delta_1$, for various $J$. We see that there is a steep increase of ${\overline{Q^{\text{noisy}}_{\text{DK}}}}$ as $\Delta_1$ approaches $\Delta_1\sim 4$, where the level crossing occurs [c.f. dotted blue curves in Fig.~\ref{Fig1} (b)]. Moreover, we observe an increase in ${\overline{Q^{\text{noisy}}_{\text{DK}}}}$ with $J$ before the level crossing. This can be understood, as the stronger $J$ is, the easier is the transition. In contrast, in the regime $\Delta_1>4$ after the level crossing, we see that ${\overline{Q^{\text{noisy}}_{\text{DK}}}}$ decreases with $J$.


Finally, for the slow Gaussian noise, ${\overline{Q^{\text{noisy}}_{\text{DK}}}}$ should be further averaged over the Gaussian-type distribution $P(J)$. We have $\langle{\overline{Q^{\text{noisy}}_{\text{DK}}}}\rangle=\int dJ P(J){\overline{Q^{\text{noisy}}_{\text{DK}}}}$. The resulting noise-averaged DK tunneling rate $\langle{\overline{Q^{\text{noisy}}_{\text{DK}}}}\rangle$ only depends on $\Delta_0$ and $\Delta_1$. As shown in Fig.~\ref{Fig3} (b), increasing $\Delta_1$ always leads to an enhanced $\langle{\overline{Q^{\text{noisy}}_{\text{DK}}}}\rangle$.

\section{Discussion and conclusion}\label{Con}

The emphasis and value of this study is a general and exactly solvable model that is capable of describing a noisy two-level quantum system. The unavoidable presence of impurities in most real-world physical systems has given
a strong motivation to the study of noisy two-level models. While usually the noisy model is considered as phenomenological in the context of condensed matter, the noisy DK model studied here is of relevance in cold-atom experiments, where the noise can be engineered~\cite{Roati2008,Billy2008}.  An optically-trapped atomic realizations of
DK model may thus serve as an ideal platform to
study the effect of time-dependent disorder in a controlled
setting by appropriate modulation of the laser beams to
mimic various noise sources~\cite{Gross2017}, thus providing an experimental counterpart of the present theoretical
study. At the same time, there also exist several other quantum simulation platforms, such as trapped ions~\cite{Blatt2012}, Rydberg atoms~\cite{Marcuzzi2017,Signoles2021} and cavity quantum electrodynamics~\cite{Sauerwein2023}, which have  displayed the capability to implement controlled disorder in the otherwise clean many-body systems. With these state-of-the-art experimental technologies, we hope the predicted results can be observed in the future experiment.

We should also bear in mind the assumptions that underlie our results. Our study is based on the two-level model and primarily focuses on transition probabilities.  In other words, our theoretical framework only considers coherent noise and
has ignored the decoherence or purity of the state after the transition is passed. To study the effects of noise on both transition probabilities and decoherence, one needs to use the generalized master equation
for the marginal system density operator~\cite{Ying2015,Zoller1981,Galperin2008,Paladino2014,Mutter2022}, which is beyond the scope of the current study.

In summary, we have explored the dynamics of the DK model in noisy environment, where the influence of the (classical) environment is modeled by telegraph noise and Gaussian noise characterized by $J\rightarrow J_{\text{noisy}}(t)$.
We analytical obtain the exact expressions for the survival probability $Q^{\text{noisy}}_{\text{DK}}$ of finding the system to remain in the initial state.
For the slow telegraph noise, we find parameter regimes where $Q^{\text{noisy}}_{\text{DK}}$ is suppressed, rather than enhanced. For the slow Gaussian noise, we find that the noise always leads to an enhanced $Q^{\text{noisy}}_{\text{DK}}$, which originate from the absorption of the noise quanta across the gap. Our study introduces a new perspective for quantum control of two-level quantum systems.

We thank Biao Wu and Ying Hu for stimulating discussions. This work was supported by the National Natural
Science Foundation of China (Nos. 12074344), the Zhejiang Provincial Natural Science Foundation (Grant Nos. LZ21A040001) and the key projects of the Natural Science Foundation of China (Grant No. 11835011).

\bibliography{Reference}

\begin{thebibliography}{49}%
\makeatletter
\providecommand \@ifxundefined [1]{%
 \@ifx{#1\undefined}
}%
\providecommand \@ifnum [1]{%
 \ifnum #1\expandafter \@firstoftwo
 \else \expandafter \@secondoftwo
 \fi
}%
\providecommand \@ifx [1]{%
 \ifx #1\expandafter \@firstoftwo
 \else \expandafter \@secondoftwo
 \fi
}%
\providecommand \natexlab [1]{#1}%
\providecommand \enquote  [1]{``#1''}%
\providecommand \bibnamefont  [1]{#1}%
\providecommand \bibfnamefont [1]{#1}%
\providecommand \citenamefont [1]{#1}%
\providecommand \href@noop [0]{\@secondoftwo}%
\providecommand \href [0]{\begingroup \@sanitize@url \@href}%
\providecommand \@href[1]{\@@startlink{#1}\@@href}%
\providecommand \@@href[1]{\endgroup#1\@@endlink}%
\providecommand \@sanitize@url [0]{\catcode `\\12\catcode `\$12\catcode
  `\&12\catcode `\#12\catcode `\^12\catcode `\_12\catcode `\%12\relax}%
\providecommand \@@startlink[1]{}%
\providecommand \@@endlink[0]{}%
\providecommand \url  [0]{\begingroup\@sanitize@url \@url }%
\providecommand \@url [1]{\endgroup\@href {#1}{\urlprefix }}%
\providecommand \urlprefix  [0]{URL }%
\providecommand \Eprint [0]{\href }%
\providecommand \doibase [0]{https://doi.org/}%
\providecommand \selectlanguage [0]{\@gobble}%
\providecommand \bibinfo  [0]{\@secondoftwo}%
\providecommand \bibfield  [0]{\@secondoftwo}%
\providecommand \translation [1]{[#1]}%
\providecommand \BibitemOpen [0]{}%
\providecommand \bibitemStop [0]{}%
\providecommand \bibitemNoStop [0]{.\EOS\space}%
\providecommand \EOS [0]{\spacefactor3000\relax}%
\providecommand \BibitemShut  [1]{\csname bibitem#1\endcsname}%
\let\auto@bib@innerbib\@empty
\bibitem [{\citenamefont {Nikitin}\ and\ \citenamefont
  {Umanskii}(1984)}]{AtomicBook}%
  \BibitemOpen
  \bibfield  {author} {\bibinfo {author} {\bibfnamefont {E.~E.}\ \bibnamefont
  {Nikitin}}\ and\ \bibinfo {author} {\bibfnamefont {S.~Y.}\ \bibnamefont
  {Umanskii}},\ }\href@noop {} {\emph {\bibinfo {title} {Theory of Slow Atomic
  Collisions, vol. 30 of Springer Series in Chemical Physics}}}\ (\bibinfo
  {publisher} {Springer Berlin Heidelberg},\ \bibinfo {year}
  {1984})\BibitemShut {NoStop}%
\bibitem [{\citenamefont {M.~N.~Leuenberger}\ and\ \citenamefont
  {Loss}(2003)}]{MagnetBook}%
  \BibitemOpen
  \bibfield  {author} {\bibinfo {author} {\bibfnamefont {F.~M.}\ \bibnamefont
  {M.~N.~Leuenberger}}\ and\ \bibinfo {author} {\bibfnamefont {D.}~\bibnamefont
  {Loss}},\ }\href@noop {} {\emph {\bibinfo {title} {Molecular Magnets Recent
  Highlights, 101-107}}}\ (\bibinfo  {publisher} {Springer},\ \bibinfo {year}
  {2003})\BibitemShut {NoStop}%
\bibitem [{\citenamefont {H\"anggi}\ \emph {et~al.}(1990)\citenamefont
  {H\"anggi}, \citenamefont {Talkner},\ and\ \citenamefont
  {Borkovec}}]{Hanggi1990}%
  \BibitemOpen
  \bibfield  {author} {\bibinfo {author} {\bibfnamefont {P.}~\bibnamefont
  {H\"anggi}}, \bibinfo {author} {\bibfnamefont {P.}~\bibnamefont {Talkner}},\
  and\ \bibinfo {author} {\bibfnamefont {M.}~\bibnamefont {Borkovec}},\
  }\bibfield  {title} {\bibinfo {title} {Reaction-rate theory: fifty years
  after kramers},\ }\href {https://doi.org/10.1103/RevModPhys.62.251}
  {\bibfield  {journal} {\bibinfo  {journal} {Rev. Mod. Phys.}\ }\textbf
  {\bibinfo {volume} {62}},\ \bibinfo {pages} {251} (\bibinfo {year}
  {1990})}\BibitemShut {NoStop}%
\bibitem [{\citenamefont {Landau}(1932)}]{Landau1932}%
  \BibitemOpen
  \bibfield  {author} {\bibinfo {author} {\bibfnamefont {L.~D.}\ \bibnamefont
  {Landau}},\ }\bibfield  {title} {\bibinfo {title} {Zur theorie der
  energieubertragung ii},\ }\href@noop {} {\bibfield  {journal} {\bibinfo
  {journal} {Phys. Z. Sowjetunion}\ }\textbf {\bibinfo {volume} {2}},\ \bibinfo
  {pages} {46} (\bibinfo {year} {1932})}\BibitemShut {NoStop}%
\bibitem [{\citenamefont {Zener}\ and\ \citenamefont
  {Fowler}(1932)}]{Zener1932}%
  \BibitemOpen
  \bibfield  {author} {\bibinfo {author} {\bibfnamefont {C.}~\bibnamefont
  {Zener}}\ and\ \bibinfo {author} {\bibfnamefont {R.~H.}\ \bibnamefont
  {Fowler}},\ }\bibfield  {title} {\bibinfo {title} {Non-adiabatic crossing of
  energy levels},\ }\href {https://doi.org/10.1098/rspa.1932.0165} {\bibfield
  {journal} {\bibinfo  {journal} {Proc. R. Soc. Lond. A}\ }\textbf {\bibinfo
  {volume} {137}},\ \bibinfo {pages} {696} (\bibinfo {year}
  {1932})}\BibitemShut {NoStop}%
\bibitem [{\citenamefont {Wu}\ and\ \citenamefont {Niu}(2000)}]{Wu2000}%
  \BibitemOpen
  \bibfield  {author} {\bibinfo {author} {\bibfnamefont {B.}~\bibnamefont
  {Wu}}\ and\ \bibinfo {author} {\bibfnamefont {Q.}~\bibnamefont {Niu}},\
  }\bibfield  {title} {\bibinfo {title} {Nonlinear landau-zener tunneling},\
  }\href {https://doi.org/10.1103/PhysRevA.61.023402} {\bibfield  {journal}
  {\bibinfo  {journal} {Phys. Rev. A}\ }\textbf {\bibinfo {volume} {61}},\
  \bibinfo {pages} {023402} (\bibinfo {year} {2000})}\BibitemShut {NoStop}%
\bibitem [{\citenamefont {Ivakhnenko}\ \emph {et~al.}(2023)\citenamefont
  {Ivakhnenko}, \citenamefont {Shevchenko},\ and\ \citenamefont
  {Nori}}]{IVAKHNENKO20231}%
  \BibitemOpen
  \bibfield  {author} {\bibinfo {author} {\bibfnamefont {O.~V.}\ \bibnamefont
  {Ivakhnenko}}, \bibinfo {author} {\bibfnamefont {S.~N.}\ \bibnamefont
  {Shevchenko}},\ and\ \bibinfo {author} {\bibfnamefont {F.}~\bibnamefont
  {Nori}},\ }\bibfield  {title} {\bibinfo {title} {Nonadiabatic
  landau–zener–stückelberg–majorana transitions, dynamics, and
  interference},\ }\href
  {https://doi.org/https://doi.org/10.1016/j.physrep.2022.10.002} {\bibfield
  {journal} {\bibinfo  {journal} {Phys. Rep.}\ }\textbf {\bibinfo {volume}
  {995}},\ \bibinfo {pages} {1} (\bibinfo {year} {2023})},\ \bibinfo {note}
  {nonadiabatic Landau-Zener-Stückelberg-Majorana transitions, dynamics, and
  interference}\BibitemShut {NoStop}%
\bibitem [{\citenamefont {Kofman}\ \emph {et~al.}(2023)\citenamefont {Kofman},
  \citenamefont {Ivakhnenko}, \citenamefont {Shevchenko},\ and\ \citenamefont
  {Nori}}]{Kofman2023}%
  \BibitemOpen
  \bibfield  {author} {\bibinfo {author} {\bibfnamefont {P.~O.}\ \bibnamefont
  {Kofman}}, \bibinfo {author} {\bibfnamefont {O.~V.}\ \bibnamefont
  {Ivakhnenko}}, \bibinfo {author} {\bibfnamefont {S.~N.}\ \bibnamefont
  {Shevchenko}},\ and\ \bibinfo {author} {\bibfnamefont {F.}~\bibnamefont
  {Nori}},\ }\bibfield  {title} {\bibinfo {title} {Majorana’s approach to
  nonadiabatic transitions validates the adiabatic-impulse approximation},\
  }\href {https://doi.org/10.1038/s41598-023-31084-y} {\bibfield  {journal}
  {\bibinfo  {journal} {Phys. Rep.}\ }\textbf {\bibinfo {volume} {13}},\
  \bibinfo {pages} {5053} (\bibinfo {year} {2023})}\BibitemShut {NoStop}%
\bibitem [{\citenamefont {Rosen}\ and\ \citenamefont {Zener}(1932)}]{RZ1932}%
  \BibitemOpen
  \bibfield  {author} {\bibinfo {author} {\bibfnamefont {N.}~\bibnamefont
  {Rosen}}\ and\ \bibinfo {author} {\bibfnamefont {C.}~\bibnamefont {Zener}},\
  }\bibfield  {title} {\bibinfo {title} {Double stern-gerlach experiment and
  related collision phenomena},\ }\href
  {https://doi.org/10.1103/PhysRev.40.502} {\bibfield  {journal} {\bibinfo
  {journal} {Phys. Rev.}\ }\textbf {\bibinfo {volume} {40}},\ \bibinfo {pages}
  {502} (\bibinfo {year} {1932})}\BibitemShut {NoStop}%
\bibitem [{\citenamefont {Hioe}(1984)}]{AEM1}%
  \BibitemOpen
  \bibfield  {author} {\bibinfo {author} {\bibfnamefont {F.~T.}\ \bibnamefont
  {Hioe}},\ }\bibfield  {title} {\bibinfo {title} {Solution of bloch equations
  involving amplitude and frequency modulations},\ }\href
  {https://doi.org/10.1103/PhysRevA.30.2100} {\bibfield  {journal} {\bibinfo
  {journal} {Phys. Rev. A}\ }\textbf {\bibinfo {volume} {30}},\ \bibinfo
  {pages} {2100} (\bibinfo {year} {1984})}\BibitemShut {NoStop}%
\bibitem [{\citenamefont {Silver}\ \emph {et~al.}(1985)\citenamefont {Silver},
  \citenamefont {Joseph},\ and\ \citenamefont {Hoult}}]{AEM2}%
  \BibitemOpen
  \bibfield  {author} {\bibinfo {author} {\bibfnamefont {M.~S.}\ \bibnamefont
  {Silver}}, \bibinfo {author} {\bibfnamefont {R.~I.}\ \bibnamefont {Joseph}},\
  and\ \bibinfo {author} {\bibfnamefont {D.~I.}\ \bibnamefont {Hoult}},\
  }\bibfield  {title} {\bibinfo {title} {Selective spin inversion in nuclear
  magnetic resonance and coherent optics through an exact solution of the
  bloch-riccati equation},\ }\href {https://doi.org/10.1103/PhysRevA.31.2753}
  {\bibfield  {journal} {\bibinfo  {journal} {Phys. Rev. A}\ }\textbf {\bibinfo
  {volume} {31}},\ \bibinfo {pages} {2753} (\bibinfo {year}
  {1985})}\BibitemShut {NoStop}%
\bibitem [{\citenamefont {Bambini}\ and\ \citenamefont {Berman}(1981)}]{BBM}%
  \BibitemOpen
  \bibfield  {author} {\bibinfo {author} {\bibfnamefont {A.}~\bibnamefont
  {Bambini}}\ and\ \bibinfo {author} {\bibfnamefont {P.~R.}\ \bibnamefont
  {Berman}},\ }\bibfield  {title} {\bibinfo {title} {Analytic solutions to the
  two-state problem for a class of coupling potentials},\ }\href
  {https://doi.org/10.1103/PhysRevA.23.2496} {\bibfield  {journal} {\bibinfo
  {journal} {Phys. Rev. A}\ }\textbf {\bibinfo {volume} {23}},\ \bibinfo
  {pages} {2496} (\bibinfo {year} {1981})}\BibitemShut {NoStop}%
\bibitem [{\citenamefont {Demkov}\ and\ \citenamefont
  {Kunike}(1969)}]{demkov1969}%
  \BibitemOpen
  \bibfield  {author} {\bibinfo {author} {\bibfnamefont {Y.~N.}\ \bibnamefont
  {Demkov}}\ and\ \bibinfo {author} {\bibfnamefont {M.}~\bibnamefont
  {Kunike}},\ }\href@noop {} {\bibfield  {journal} {\bibinfo  {journal} {Vestn.
  Leningr. Univ., Ser. 4, Fiz. Khim.}\ }\textbf {\bibinfo {volume} {16}},\
  \bibinfo {pages} {39} (\bibinfo {year} {1969})}\BibitemShut {NoStop}%
\bibitem [{\citenamefont {Hioe}\ and\ \citenamefont {Carroll}(1985)}]{DKM1}%
  \BibitemOpen
  \bibfield  {author} {\bibinfo {author} {\bibfnamefont {F.~T.}\ \bibnamefont
  {Hioe}}\ and\ \bibinfo {author} {\bibfnamefont {C.~E.}\ \bibnamefont
  {Carroll}},\ }\bibfield  {title} {\bibinfo {title} {Two-state problems
  involving arbitrary amplitude and frequency modulations},\ }\href
  {https://doi.org/10.1103/PhysRevA.32.1541} {\bibfield  {journal} {\bibinfo
  {journal} {Phys. Rev. A}\ }\textbf {\bibinfo {volume} {32}},\ \bibinfo
  {pages} {1541} (\bibinfo {year} {1985})}\BibitemShut {NoStop}%
\bibitem [{\citenamefont {Zakrzewski}(1985)}]{DKM2}%
  \BibitemOpen
  \bibfield  {author} {\bibinfo {author} {\bibfnamefont {J.}~\bibnamefont
  {Zakrzewski}},\ }\bibfield  {title} {\bibinfo {title} {Analytic solutions of
  the two-state problem for a class of chirped pulses},\ }\href
  {https://doi.org/10.1103/PhysRevA.32.3748} {\bibfield  {journal} {\bibinfo
  {journal} {Phys. Rev. A}\ }\textbf {\bibinfo {volume} {32}},\ \bibinfo
  {pages} {3748} (\bibinfo {year} {1985})}\BibitemShut {NoStop}%
\bibitem [{\citenamefont {Suominen}\ and\ \citenamefont
  {Garraway}(1992)}]{DKM3}%
  \BibitemOpen
  \bibfield  {author} {\bibinfo {author} {\bibfnamefont {K.-A.}\ \bibnamefont
  {Suominen}}\ and\ \bibinfo {author} {\bibfnamefont {B.~M.}\ \bibnamefont
  {Garraway}},\ }\bibfield  {title} {\bibinfo {title} {Population transfer in a
  level-crossing model with two time scales},\ }\href
  {https://doi.org/10.1103/PhysRevA.45.374} {\bibfield  {journal} {\bibinfo
  {journal} {Phys. Rev. A}\ }\textbf {\bibinfo {volume} {45}},\ \bibinfo
  {pages} {374} (\bibinfo {year} {1992})}\BibitemShut {NoStop}%
\bibitem [{\citenamefont {Lacour}\ \emph {et~al.}(2007)\citenamefont {Lacour},
  \citenamefont {Gu\'erin}, \citenamefont {Yatsenko}, \citenamefont {Vitanov},\
  and\ \citenamefont {Jauslin}}]{Lacour2007}%
  \BibitemOpen
  \bibfield  {author} {\bibinfo {author} {\bibfnamefont {X.}~\bibnamefont
  {Lacour}}, \bibinfo {author} {\bibfnamefont {S.}~\bibnamefont {Gu\'erin}},
  \bibinfo {author} {\bibfnamefont {L.~P.}\ \bibnamefont {Yatsenko}}, \bibinfo
  {author} {\bibfnamefont {N.~V.}\ \bibnamefont {Vitanov}},\ and\ \bibinfo
  {author} {\bibfnamefont {H.~R.}\ \bibnamefont {Jauslin}},\ }\bibfield
  {title} {\bibinfo {title} {Uniform analytic description of dephasing effects
  in two-state transitions},\ }\href
  {https://doi.org/10.1103/PhysRevA.75.033417} {\bibfield  {journal} {\bibinfo
  {journal} {Phys. Rev. A}\ }\textbf {\bibinfo {volume} {75}},\ \bibinfo
  {pages} {033417} (\bibinfo {year} {2007})}\BibitemShut {NoStop}%
\bibitem [{\citenamefont {Simeonov}\ and\ \citenamefont
  {Vitanov}(2014)}]{Simeonov2014}%
  \BibitemOpen
  \bibfield  {author} {\bibinfo {author} {\bibfnamefont {L.~S.}\ \bibnamefont
  {Simeonov}}\ and\ \bibinfo {author} {\bibfnamefont {N.~V.}\ \bibnamefont
  {Vitanov}},\ }\bibfield  {title} {\bibinfo {title} {Exactly solvable
  two-state quantum model for a pulse of hyperbolic-tangent shape},\ }\href
  {https://doi.org/10.1103/PhysRevA.89.043411} {\bibfield  {journal} {\bibinfo
  {journal} {Phys. Rev. A}\ }\textbf {\bibinfo {volume} {89}},\ \bibinfo
  {pages} {043411} (\bibinfo {year} {2014})}\BibitemShut {NoStop}%
\bibitem [{\citenamefont {Sillanp\"a\"a}\ \emph {et~al.}(2006)\citenamefont
  {Sillanp\"a\"a}, \citenamefont {Lehtinen}, \citenamefont {Paila},
  \citenamefont {Makhlin},\ and\ \citenamefont {Hakonen}}]{Sillanpaa2006}%
  \BibitemOpen
  \bibfield  {author} {\bibinfo {author} {\bibfnamefont {M.}~\bibnamefont
  {Sillanp\"a\"a}}, \bibinfo {author} {\bibfnamefont {T.}~\bibnamefont
  {Lehtinen}}, \bibinfo {author} {\bibfnamefont {A.}~\bibnamefont {Paila}},
  \bibinfo {author} {\bibfnamefont {Y.}~\bibnamefont {Makhlin}},\ and\ \bibinfo
  {author} {\bibfnamefont {P.}~\bibnamefont {Hakonen}},\ }\bibfield  {title}
  {\bibinfo {title} {Continuous-time monitoring of landau-zener interference in
  a cooper-pair box},\ }\href {https://doi.org/10.1103/PhysRevLett.96.187002}
  {\bibfield  {journal} {\bibinfo  {journal} {Phys. Rev. Lett.}\ }\textbf
  {\bibinfo {volume} {96}},\ \bibinfo {pages} {187002} (\bibinfo {year}
  {2006})}\BibitemShut {NoStop}%
\bibitem [{\citenamefont {Crosson}\ and\ \citenamefont
  {Lidar}(2021)}]{Crosson2021}%
  \BibitemOpen
  \bibfield  {author} {\bibinfo {author} {\bibfnamefont {E.~J.}\ \bibnamefont
  {Crosson}}\ and\ \bibinfo {author} {\bibfnamefont {D.~A.}\ \bibnamefont
  {Lidar}},\ }\bibfield  {title} {\bibinfo {title} {Prospects for quantum
  enhancement with diabatic quantum annealing},\ }\href
  {https://www.nature.com/articles/s42254-021-00313-6} {\bibfield  {journal}
  {\bibinfo  {journal} {Nat. Rew. Phys.}\ }\textbf {\bibinfo {volume} {3}},\
  \bibinfo {pages} {466} (\bibinfo {year} {2021})}\BibitemShut {NoStop}%
\bibitem [{\citenamefont {Preskill}(2018)}]{Preskill2018}%
  \BibitemOpen
  \bibfield  {author} {\bibinfo {author} {\bibfnamefont {J.}~\bibnamefont
  {Preskill}},\ }\bibfield  {title} {\bibinfo {title} {Quantum {C}omputing in
  the {NISQ} era and beyond},\ }\href
  {https://doi.org/10.22331/q-2018-08-06-79} {\bibfield  {journal} {\bibinfo
  {journal} {{Quantum}}\ }\textbf {\bibinfo {volume} {2}},\ \bibinfo {pages}
  {79} (\bibinfo {year} {2018})}\BibitemShut {NoStop}%
\bibitem [{\citenamefont {Cheng}\ \emph {et~al.}(2023)\citenamefont {Cheng},
  \citenamefont {Deng}, \citenamefont {Gu}, \citenamefont {He}, \citenamefont
  {Hu}, \citenamefont {Huang}, \citenamefont {Li}, \citenamefont {Lin},
  \citenamefont {Lu}, \citenamefont {Lu}, \citenamefont {Qiu}, \citenamefont
  {Wang}, \citenamefont {Xin}, \citenamefont {Yu}, \citenamefont {Yung},
  \citenamefont {Zeng}, \citenamefont {Zhang}, \citenamefont {Zhong},
  \citenamefont {Peng}, \citenamefont {Nori},\ and\ \citenamefont
  {Yu}}]{Cheng2023}%
  \BibitemOpen
  \bibfield  {author} {\bibinfo {author} {\bibfnamefont {B.}~\bibnamefont
  {Cheng}}, \bibinfo {author} {\bibfnamefont {X.-H.}\ \bibnamefont {Deng}},
  \bibinfo {author} {\bibfnamefont {X.}~\bibnamefont {Gu}}, \bibinfo {author}
  {\bibfnamefont {Y.}~\bibnamefont {He}}, \bibinfo {author} {\bibfnamefont
  {G.}~\bibnamefont {Hu}}, \bibinfo {author} {\bibfnamefont {P.}~\bibnamefont
  {Huang}}, \bibinfo {author} {\bibfnamefont {J.}~\bibnamefont {Li}}, \bibinfo
  {author} {\bibfnamefont {B.-C.}\ \bibnamefont {Lin}}, \bibinfo {author}
  {\bibfnamefont {D.}~\bibnamefont {Lu}}, \bibinfo {author} {\bibfnamefont
  {Y.}~\bibnamefont {Lu}}, \bibinfo {author} {\bibfnamefont {C.}~\bibnamefont
  {Qiu}}, \bibinfo {author} {\bibfnamefont {H.}~\bibnamefont {Wang}}, \bibinfo
  {author} {\bibfnamefont {T.}~\bibnamefont {Xin}}, \bibinfo {author}
  {\bibfnamefont {S.}~\bibnamefont {Yu}}, \bibinfo {author} {\bibfnamefont
  {M.-H.}\ \bibnamefont {Yung}}, \bibinfo {author} {\bibfnamefont
  {J.}~\bibnamefont {Zeng}}, \bibinfo {author} {\bibfnamefont {S.}~\bibnamefont
  {Zhang}}, \bibinfo {author} {\bibfnamefont {Y.}~\bibnamefont {Zhong}},
  \bibinfo {author} {\bibfnamefont {X.}~\bibnamefont {Peng}}, \bibinfo {author}
  {\bibfnamefont {F.}~\bibnamefont {Nori}},\ and\ \bibinfo {author}
  {\bibfnamefont {D.}~\bibnamefont {Yu}},\ }\bibfield  {title} {\bibinfo
  {title} {Noisy intermediate-scale quantum computers},\ }\href
  {https://doi.org/10.1007/s11467-022-1249-z} {\bibfield  {journal} {\bibinfo
  {journal} {Front. Phys.}\ }\textbf {\bibinfo {volume} {18}},\ \bibinfo
  {pages} {21308} (\bibinfo {year} {2023})}\BibitemShut {NoStop}%
\bibitem [{\citenamefont {Krantz}\ \emph {et~al.}(2019)\citenamefont {Krantz},
  \citenamefont {Kjaergaard}, \citenamefont {Yan}, \citenamefont {Orlando},
  \citenamefont {Gustavsson},\ and\ \citenamefont {Oliver}}]{YanFei2019}%
  \BibitemOpen
  \bibfield  {author} {\bibinfo {author} {\bibfnamefont {P.}~\bibnamefont
  {Krantz}}, \bibinfo {author} {\bibfnamefont {M.}~\bibnamefont {Kjaergaard}},
  \bibinfo {author} {\bibfnamefont {F.}~\bibnamefont {Yan}}, \bibinfo {author}
  {\bibfnamefont {T.~P.}\ \bibnamefont {Orlando}}, \bibinfo {author}
  {\bibfnamefont {S.}~\bibnamefont {Gustavsson}},\ and\ \bibinfo {author}
  {\bibfnamefont {W.~D.}\ \bibnamefont {Oliver}},\ }\bibfield  {title}
  {\bibinfo {title} {{A quantum engineer's guide to superconducting qubits}},\
  }\href {https://doi.org/10.1063/1.5089550} {\bibfield  {journal} {\bibinfo
  {journal} {Appl. Phys. Rev.}\ }\textbf {\bibinfo {volume} {6}},\ \bibinfo
  {pages} {021318} (\bibinfo {year} {2019})}\BibitemShut {NoStop}%
\bibitem [{\citenamefont {Kayanuma}(1987)}]{Kayanuma1987}%
  \BibitemOpen
  \bibfield  {author} {\bibinfo {author} {\bibfnamefont {Y.}~\bibnamefont
  {Kayanuma}},\ }\bibfield  {title} {\bibinfo {title} {Population inversion in
  optical adiabatic rapid passage with phase relaxation},\ }\href
  {https://doi.org/10.1103/PhysRevLett.58.1934} {\bibfield  {journal} {\bibinfo
   {journal} {Phys. Rev. Lett.}\ }\textbf {\bibinfo {volume} {58}},\ \bibinfo
  {pages} {1934} (\bibinfo {year} {1987})}\BibitemShut {NoStop}%
\bibitem [{\citenamefont {Kayanuma}\ and\ \citenamefont
  {Nakayama}(1998)}]{Kayanuma1998}%
  \BibitemOpen
  \bibfield  {author} {\bibinfo {author} {\bibfnamefont {Y.}~\bibnamefont
  {Kayanuma}}\ and\ \bibinfo {author} {\bibfnamefont {H.}~\bibnamefont
  {Nakayama}},\ }\bibfield  {title} {\bibinfo {title} {Nonadiabatic transition
  at a level crossing with dissipation},\ }\href
  {https://doi.org/10.1103/PhysRevB.57.13099} {\bibfield  {journal} {\bibinfo
  {journal} {Phys. Rev. B}\ }\textbf {\bibinfo {volume} {57}},\ \bibinfo
  {pages} {13099} (\bibinfo {year} {1998})}\BibitemShut {NoStop}%
\bibitem [{\citenamefont {Ao}\ and\ \citenamefont {Rammer}(1989)}]{Ao1989}%
  \BibitemOpen
  \bibfield  {author} {\bibinfo {author} {\bibfnamefont {P.}~\bibnamefont
  {Ao}}\ and\ \bibinfo {author} {\bibfnamefont {J.}~\bibnamefont {Rammer}},\
  }\bibfield  {title} {\bibinfo {title} {Influence of dissipation on the
  landau-zener transition},\ }\href
  {https://doi.org/10.1103/PhysRevLett.62.3004} {\bibfield  {journal} {\bibinfo
   {journal} {Phys. Rev. Lett.}\ }\textbf {\bibinfo {volume} {62}},\ \bibinfo
  {pages} {3004} (\bibinfo {year} {1989})}\BibitemShut {NoStop}%
\bibitem [{\citenamefont {Ao}\ and\ \citenamefont {Rammer}(1991)}]{Ao1991}%
  \BibitemOpen
  \bibfield  {author} {\bibinfo {author} {\bibfnamefont {P.}~\bibnamefont
  {Ao}}\ and\ \bibinfo {author} {\bibfnamefont {J.}~\bibnamefont {Rammer}},\
  }\bibfield  {title} {\bibinfo {title} {Quantum dynamics of a two-state system
  in a dissipative environment},\ }\href
  {https://doi.org/10.1103/PhysRevB.43.5397} {\bibfield  {journal} {\bibinfo
  {journal} {Phys. Rev. B}\ }\textbf {\bibinfo {volume} {43}},\ \bibinfo
  {pages} {5397} (\bibinfo {year} {1991})}\BibitemShut {NoStop}%
\bibitem [{\citenamefont {Pokrovsky}\ and\ \citenamefont
  {Sinitsyn}(2003)}]{Pokrovsky2003}%
  \BibitemOpen
  \bibfield  {author} {\bibinfo {author} {\bibfnamefont {V.~L.}\ \bibnamefont
  {Pokrovsky}}\ and\ \bibinfo {author} {\bibfnamefont {N.~A.}\ \bibnamefont
  {Sinitsyn}},\ }\bibfield  {title} {\bibinfo {title} {Fast noise in the
  landau-zener theory},\ }\href {https://doi.org/10.1103/PhysRevB.67.144303}
  {\bibfield  {journal} {\bibinfo  {journal} {Phys. Rev. B}\ }\textbf {\bibinfo
  {volume} {67}},\ \bibinfo {pages} {144303} (\bibinfo {year}
  {2003})}\BibitemShut {NoStop}%
\bibitem [{\citenamefont {Pokrovsky}\ and\ \citenamefont
  {Sun}(2007)}]{Pokrovsky2007}%
  \BibitemOpen
  \bibfield  {author} {\bibinfo {author} {\bibfnamefont {V.~L.}\ \bibnamefont
  {Pokrovsky}}\ and\ \bibinfo {author} {\bibfnamefont {D.}~\bibnamefont
  {Sun}},\ }\bibfield  {title} {\bibinfo {title} {Fast quantum noise in the
  landau-zener transition},\ }\href
  {https://doi.org/10.1103/PhysRevB.76.024310} {\bibfield  {journal} {\bibinfo
  {journal} {Phys. Rev. B}\ }\textbf {\bibinfo {volume} {76}},\ \bibinfo
  {pages} {024310} (\bibinfo {year} {2007})}\BibitemShut {NoStop}%
\bibitem [{\citenamefont {Nalbach}\ and\ \citenamefont
  {Thorwart}(2009)}]{Nalbach2009}%
  \BibitemOpen
  \bibfield  {author} {\bibinfo {author} {\bibfnamefont {P.}~\bibnamefont
  {Nalbach}}\ and\ \bibinfo {author} {\bibfnamefont {M.}~\bibnamefont
  {Thorwart}},\ }\bibfield  {title} {\bibinfo {title} {Landau-zener transitions
  in a dissipative environment: Numerically exact results},\ }\href
  {https://doi.org/10.1103/PhysRevLett.103.220401} {\bibfield  {journal}
  {\bibinfo  {journal} {Phys. Rev. Lett.}\ }\textbf {\bibinfo {volume} {103}},\
  \bibinfo {pages} {220401} (\bibinfo {year} {2009})}\BibitemShut {NoStop}%
\bibitem [{\citenamefont {Luo}\ and\ \citenamefont {Raikh}(2017)}]{Luo2017}%
  \BibitemOpen
  \bibfield  {author} {\bibinfo {author} {\bibfnamefont {Z.-X.}\ \bibnamefont
  {Luo}}\ and\ \bibinfo {author} {\bibfnamefont {M.~E.}\ \bibnamefont
  {Raikh}},\ }\bibfield  {title} {\bibinfo {title} {Landau-zener transition
  driven by slow noise},\ }\href {https://doi.org/10.1103/PhysRevB.95.064305}
  {\bibfield  {journal} {\bibinfo  {journal} {Phys. Rev. B}\ }\textbf {\bibinfo
  {volume} {95}},\ \bibinfo {pages} {064305} (\bibinfo {year}
  {2017})}\BibitemShut {NoStop}%
\bibitem [{\citenamefont {Malla}\ \emph {et~al.}(2017)\citenamefont {Malla},
  \citenamefont {Mishchenko},\ and\ \citenamefont {Raikh}}]{Malla2017}%
  \BibitemOpen
  \bibfield  {author} {\bibinfo {author} {\bibfnamefont {R.~K.}\ \bibnamefont
  {Malla}}, \bibinfo {author} {\bibfnamefont {E.~G.}\ \bibnamefont
  {Mishchenko}},\ and\ \bibinfo {author} {\bibfnamefont {M.~E.}\ \bibnamefont
  {Raikh}},\ }\bibfield  {title} {\bibinfo {title} {Suppression of the
  landau-zener transition probability by weak classical noise},\ }\href
  {https://doi.org/10.1103/PhysRevB.96.075419} {\bibfield  {journal} {\bibinfo
  {journal} {Phys. Rev. B}\ }\textbf {\bibinfo {volume} {96}},\ \bibinfo
  {pages} {075419} (\bibinfo {year} {2017})}\BibitemShut {NoStop}%
\bibitem [{\citenamefont {Harris}\ \emph {et~al.}(2008)\citenamefont {Harris},
  \citenamefont {Johnson}, \citenamefont {Han}, \citenamefont {Berkley},
  \citenamefont {Johansson}, \citenamefont {Bunyk}, \citenamefont {Ladizinsky},
  \citenamefont {Govorkov}, \citenamefont {Thom}, \citenamefont {Uchaikin},
  \citenamefont {Bumble}, \citenamefont {Fung}, \citenamefont {Kaul},
  \citenamefont {Kleinsasser}, \citenamefont {Amin},\ and\ \citenamefont
  {Averin}}]{Harris2008}%
  \BibitemOpen
  \bibfield  {author} {\bibinfo {author} {\bibfnamefont {R.}~\bibnamefont
  {Harris}}, \bibinfo {author} {\bibfnamefont {M.~W.}\ \bibnamefont {Johnson}},
  \bibinfo {author} {\bibfnamefont {S.}~\bibnamefont {Han}}, \bibinfo {author}
  {\bibfnamefont {A.~J.}\ \bibnamefont {Berkley}}, \bibinfo {author}
  {\bibfnamefont {J.}~\bibnamefont {Johansson}}, \bibinfo {author}
  {\bibfnamefont {P.}~\bibnamefont {Bunyk}}, \bibinfo {author} {\bibfnamefont
  {E.}~\bibnamefont {Ladizinsky}}, \bibinfo {author} {\bibfnamefont
  {S.}~\bibnamefont {Govorkov}}, \bibinfo {author} {\bibfnamefont {M.~C.}\
  \bibnamefont {Thom}}, \bibinfo {author} {\bibfnamefont {S.}~\bibnamefont
  {Uchaikin}}, \bibinfo {author} {\bibfnamefont {B.}~\bibnamefont {Bumble}},
  \bibinfo {author} {\bibfnamefont {A.}~\bibnamefont {Fung}}, \bibinfo {author}
  {\bibfnamefont {A.}~\bibnamefont {Kaul}}, \bibinfo {author} {\bibfnamefont
  {A.}~\bibnamefont {Kleinsasser}}, \bibinfo {author} {\bibfnamefont
  {M.~H.~S.}\ \bibnamefont {Amin}},\ and\ \bibinfo {author} {\bibfnamefont
  {D.~V.}\ \bibnamefont {Averin}},\ }\bibfield  {title} {\bibinfo {title}
  {Probing noise in flux qubits via macroscopic resonant tunneling},\ }\href
  {https://doi.org/10.1103/PhysRevLett.101.117003} {\bibfield  {journal}
  {\bibinfo  {journal} {Phys. Rev. Lett.}\ }\textbf {\bibinfo {volume} {101}},\
  \bibinfo {pages} {117003} (\bibinfo {year} {2008})}\BibitemShut {NoStop}%
\bibitem [{\citenamefont {Quintana}\ \emph {et~al.}(2017)\citenamefont
  {Quintana}, \citenamefont {Chen}, \citenamefont {Sank}, \citenamefont
  {Petukhov}, \citenamefont {White}, \citenamefont {Kafri}, \citenamefont
  {Chiaro}, \citenamefont {Megrant}, \citenamefont {Barends}, \citenamefont
  {Campbell}, \citenamefont {Chen}, \citenamefont {Dunsworth}, \citenamefont
  {Fowler}, \citenamefont {Graff}, \citenamefont {Jeffrey}, \citenamefont
  {Kelly}, \citenamefont {Lucero}, \citenamefont {Mutus}, \citenamefont
  {Neeley}, \citenamefont {Neill}, \citenamefont {O'Malley}, \citenamefont
  {Roushan}, \citenamefont {Shabani}, \citenamefont {Smelyanskiy},
  \citenamefont {Vainsencher}, \citenamefont {Wenner}, \citenamefont {Neven},\
  and\ \citenamefont {Martinis}}]{Quintana2017}%
  \BibitemOpen
  \bibfield  {author} {\bibinfo {author} {\bibfnamefont {C.~M.}\ \bibnamefont
  {Quintana}}, \bibinfo {author} {\bibfnamefont {Y.}~\bibnamefont {Chen}},
  \bibinfo {author} {\bibfnamefont {D.}~\bibnamefont {Sank}}, \bibinfo {author}
  {\bibfnamefont {A.~G.}\ \bibnamefont {Petukhov}}, \bibinfo {author}
  {\bibfnamefont {T.~C.}\ \bibnamefont {White}}, \bibinfo {author}
  {\bibfnamefont {D.}~\bibnamefont {Kafri}}, \bibinfo {author} {\bibfnamefont
  {B.}~\bibnamefont {Chiaro}}, \bibinfo {author} {\bibfnamefont
  {A.}~\bibnamefont {Megrant}}, \bibinfo {author} {\bibfnamefont
  {R.}~\bibnamefont {Barends}}, \bibinfo {author} {\bibfnamefont
  {B.}~\bibnamefont {Campbell}}, \bibinfo {author} {\bibfnamefont
  {Z.}~\bibnamefont {Chen}}, \bibinfo {author} {\bibfnamefont {A.}~\bibnamefont
  {Dunsworth}}, \bibinfo {author} {\bibfnamefont {A.~G.}\ \bibnamefont
  {Fowler}}, \bibinfo {author} {\bibfnamefont {R.}~\bibnamefont {Graff}},
  \bibinfo {author} {\bibfnamefont {E.}~\bibnamefont {Jeffrey}}, \bibinfo
  {author} {\bibfnamefont {J.}~\bibnamefont {Kelly}}, \bibinfo {author}
  {\bibfnamefont {E.}~\bibnamefont {Lucero}}, \bibinfo {author} {\bibfnamefont
  {J.~Y.}\ \bibnamefont {Mutus}}, \bibinfo {author} {\bibfnamefont
  {M.}~\bibnamefont {Neeley}}, \bibinfo {author} {\bibfnamefont
  {C.}~\bibnamefont {Neill}}, \bibinfo {author} {\bibfnamefont {P.~J.~J.}\
  \bibnamefont {O'Malley}}, \bibinfo {author} {\bibfnamefont {P.}~\bibnamefont
  {Roushan}}, \bibinfo {author} {\bibfnamefont {A.}~\bibnamefont {Shabani}},
  \bibinfo {author} {\bibfnamefont {V.~N.}\ \bibnamefont {Smelyanskiy}},
  \bibinfo {author} {\bibfnamefont {A.}~\bibnamefont {Vainsencher}}, \bibinfo
  {author} {\bibfnamefont {J.}~\bibnamefont {Wenner}}, \bibinfo {author}
  {\bibfnamefont {H.}~\bibnamefont {Neven}},\ and\ \bibinfo {author}
  {\bibfnamefont {J.~M.}\ \bibnamefont {Martinis}},\ }\bibfield  {title}
  {\bibinfo {title} {Observation of classical-quantum crossover of $1/f$ flux
  noise and its paramagnetic temperature dependence},\ }\href
  {https://doi.org/10.1103/PhysRevLett.118.057702} {\bibfield  {journal}
  {\bibinfo  {journal} {Phys. Rev. Lett.}\ }\textbf {\bibinfo {volume} {118}},\
  \bibinfo {pages} {057702} (\bibinfo {year} {2017})}\BibitemShut {NoStop}%
\bibitem [{\citenamefont {Rower}\ \emph {et~al.}(2023)\citenamefont {Rower},
  \citenamefont {Ateshian}, \citenamefont {Li}, \citenamefont {Hays},
  \citenamefont {Bluvstein}, \citenamefont {Ding}, \citenamefont {Kannan},
  \citenamefont {Almanakly}, \citenamefont {Braum\"uller}, \citenamefont {Kim},
  \citenamefont {Melville}, \citenamefont {Niedzielski}, \citenamefont
  {Schwartz}, \citenamefont {Yoder}, \citenamefont {Orlando}, \citenamefont
  {Wang}, \citenamefont {Gustavsson}, \citenamefont {Grover}, \citenamefont
  {Serniak}, \citenamefont {Comin},\ and\ \citenamefont {Oliver}}]{Rower2023}%
  \BibitemOpen
  \bibfield  {author} {\bibinfo {author} {\bibfnamefont {D.~A.}\ \bibnamefont
  {Rower}}, \bibinfo {author} {\bibfnamefont {L.}~\bibnamefont {Ateshian}},
  \bibinfo {author} {\bibfnamefont {L.~H.}\ \bibnamefont {Li}}, \bibinfo
  {author} {\bibfnamefont {M.}~\bibnamefont {Hays}}, \bibinfo {author}
  {\bibfnamefont {D.}~\bibnamefont {Bluvstein}}, \bibinfo {author}
  {\bibfnamefont {L.}~\bibnamefont {Ding}}, \bibinfo {author} {\bibfnamefont
  {B.}~\bibnamefont {Kannan}}, \bibinfo {author} {\bibfnamefont
  {A.}~\bibnamefont {Almanakly}}, \bibinfo {author} {\bibfnamefont
  {J.}~\bibnamefont {Braum\"uller}}, \bibinfo {author} {\bibfnamefont {D.~K.}\
  \bibnamefont {Kim}}, \bibinfo {author} {\bibfnamefont {A.}~\bibnamefont
  {Melville}}, \bibinfo {author} {\bibfnamefont {B.~M.}\ \bibnamefont
  {Niedzielski}}, \bibinfo {author} {\bibfnamefont {M.~E.}\ \bibnamefont
  {Schwartz}}, \bibinfo {author} {\bibfnamefont {J.~L.}\ \bibnamefont {Yoder}},
  \bibinfo {author} {\bibfnamefont {T.~P.}\ \bibnamefont {Orlando}}, \bibinfo
  {author} {\bibfnamefont {J.-J.}\ \bibnamefont {Wang}}, \bibinfo {author}
  {\bibfnamefont {S.}~\bibnamefont {Gustavsson}}, \bibinfo {author}
  {\bibfnamefont {J.~A.}\ \bibnamefont {Grover}}, \bibinfo {author}
  {\bibfnamefont {K.}~\bibnamefont {Serniak}}, \bibinfo {author} {\bibfnamefont
  {R.}~\bibnamefont {Comin}},\ and\ \bibinfo {author} {\bibfnamefont {W.~D.}\
  \bibnamefont {Oliver}},\ }\bibfield  {title} {\bibinfo {title} {Evolution of
  $1/f$ flux noise in superconducting qubits with weak magnetic fields},\
  }\href {https://doi.org/10.1103/PhysRevLett.130.220602} {\bibfield  {journal}
  {\bibinfo  {journal} {Phys. Rev. Lett.}\ }\textbf {\bibinfo {volume} {130}},\
  \bibinfo {pages} {220602} (\bibinfo {year} {2023})}\BibitemShut {NoStop}%
\bibitem [{\citenamefont {Dai}\ \emph {et~al.}(2022)\citenamefont {Dai},
  \citenamefont {Trappen}, \citenamefont {Chen}, \citenamefont {Melanson},
  \citenamefont {Yurtalan}, \citenamefont {Tennant}, \citenamefont {Martinez},
  \citenamefont {Tang}, \citenamefont {Mozgunov}, \citenamefont {Gibson},
  \citenamefont {Grover}, \citenamefont {Disseler}, \citenamefont {Basham},
  \citenamefont {Novikov}, \citenamefont {Das}, \citenamefont {Melville},
  \citenamefont {Niedzielski}, \citenamefont {Hirjibehedin}, \citenamefont
  {Serniak}, \citenamefont {Weber}, \citenamefont {Yoder}, \citenamefont
  {Oliver}, \citenamefont {Zick}, \citenamefont {Lidar},\ and\ \citenamefont
  {Lupascu}}]{dai2022dissipative}%
  \BibitemOpen
  \bibfield  {author} {\bibinfo {author} {\bibfnamefont {X.}~\bibnamefont
  {Dai}}, \bibinfo {author} {\bibfnamefont {R.}~\bibnamefont {Trappen}},
  \bibinfo {author} {\bibfnamefont {H.}~\bibnamefont {Chen}}, \bibinfo {author}
  {\bibfnamefont {D.}~\bibnamefont {Melanson}}, \bibinfo {author}
  {\bibfnamefont {M.~A.}\ \bibnamefont {Yurtalan}}, \bibinfo {author}
  {\bibfnamefont {D.~M.}\ \bibnamefont {Tennant}}, \bibinfo {author}
  {\bibfnamefont {A.~J.}\ \bibnamefont {Martinez}}, \bibinfo {author}
  {\bibfnamefont {Y.}~\bibnamefont {Tang}}, \bibinfo {author} {\bibfnamefont
  {E.}~\bibnamefont {Mozgunov}}, \bibinfo {author} {\bibfnamefont
  {J.}~\bibnamefont {Gibson}}, \bibinfo {author} {\bibfnamefont {J.~A.}\
  \bibnamefont {Grover}}, \bibinfo {author} {\bibfnamefont {S.~M.}\
  \bibnamefont {Disseler}}, \bibinfo {author} {\bibfnamefont {J.~I.}\
  \bibnamefont {Basham}}, \bibinfo {author} {\bibfnamefont {S.}~\bibnamefont
  {Novikov}}, \bibinfo {author} {\bibfnamefont {R.}~\bibnamefont {Das}},
  \bibinfo {author} {\bibfnamefont {A.~J.}\ \bibnamefont {Melville}}, \bibinfo
  {author} {\bibfnamefont {B.~M.}\ \bibnamefont {Niedzielski}}, \bibinfo
  {author} {\bibfnamefont {C.~F.}\ \bibnamefont {Hirjibehedin}}, \bibinfo
  {author} {\bibfnamefont {K.}~\bibnamefont {Serniak}}, \bibinfo {author}
  {\bibfnamefont {S.~J.}\ \bibnamefont {Weber}}, \bibinfo {author}
  {\bibfnamefont {J.~L.}\ \bibnamefont {Yoder}}, \bibinfo {author}
  {\bibfnamefont {W.~D.}\ \bibnamefont {Oliver}}, \bibinfo {author}
  {\bibfnamefont {K.~M.}\ \bibnamefont {Zick}}, \bibinfo {author}
  {\bibfnamefont {D.~A.}\ \bibnamefont {Lidar}},\ and\ \bibinfo {author}
  {\bibfnamefont {A.}~\bibnamefont {Lupascu}},\ }\href@noop {} {\bibinfo
  {title} {Dissipative landau-zener tunneling: crossover from weak to strong
  environment coupling}} (\bibinfo {year} {2022}),\ \Eprint
  {https://arxiv.org/abs/2207.02017} {arXiv:2207.02017 [quant-ph]} \BibitemShut
  {NoStop}%
\bibitem [{\citenamefont {Hu}\ \emph {et~al.}(2015)\citenamefont {Hu},
  \citenamefont {Cai}, \citenamefont {Baranov},\ and\ \citenamefont
  {Zoller}}]{Ying2015}%
  \BibitemOpen
  \bibfield  {author} {\bibinfo {author} {\bibfnamefont {Y.}~\bibnamefont
  {Hu}}, \bibinfo {author} {\bibfnamefont {Z.}~\bibnamefont {Cai}}, \bibinfo
  {author} {\bibfnamefont {M.~A.}\ \bibnamefont {Baranov}},\ and\ \bibinfo
  {author} {\bibfnamefont {P.}~\bibnamefont {Zoller}},\ }\bibfield  {title}
  {\bibinfo {title} {Majorana fermions in noisy kitaev wires},\ }\href
  {https://doi.org/10.1103/PhysRevB.92.165118} {\bibfield  {journal} {\bibinfo
  {journal} {Phys. Rev. B}\ }\textbf {\bibinfo {volume} {92}},\ \bibinfo
  {pages} {165118} (\bibinfo {year} {2015})}\BibitemShut {NoStop}%
\bibitem [{\citenamefont {Zoller}\ \emph {et~al.}(1981)\citenamefont {Zoller},
  \citenamefont {Alber},\ and\ \citenamefont {Salvador}}]{Zoller1981}%
  \BibitemOpen
  \bibfield  {author} {\bibinfo {author} {\bibfnamefont {P.}~\bibnamefont
  {Zoller}}, \bibinfo {author} {\bibfnamefont {G.}~\bibnamefont {Alber}},\ and\
  \bibinfo {author} {\bibfnamefont {R.}~\bibnamefont {Salvador}},\ }\bibfield
  {title} {\bibinfo {title} {ac stark splitting in intense stochastic driving
  fields with gaussian statistics and non-lorentzian line shape},\ }\href
  {https://doi.org/10.1103/PhysRevA.24.398} {\bibfield  {journal} {\bibinfo
  {journal} {Phys. Rev. A}\ }\textbf {\bibinfo {volume} {24}},\ \bibinfo
  {pages} {398} (\bibinfo {year} {1981})}\BibitemShut {NoStop}%
\bibitem [{\citenamefont {Kayanuma}(1985)}]{Kayanuma1985}%
  \BibitemOpen
  \bibfield  {author} {\bibinfo {author} {\bibfnamefont {Y.}~\bibnamefont
  {Kayanuma}},\ }\bibfield  {title} {\bibinfo {title} {Stochastic theory for
  nonadiabatic level crossing with fluctuating off-diagonal coupling},\ }\href
  {https://doi.org/10.1143/JPSJ.54.2037} {\bibfield  {journal} {\bibinfo
  {journal} {J. Phys. Soc. Japan}\ }\textbf {\bibinfo {volume} {54}},\ \bibinfo
  {pages} {2037} (\bibinfo {year} {1985})},\ \Eprint
  {https://arxiv.org/abs/https://doi.org/10.1143/JPSJ.54.2037}
  {https://doi.org/10.1143/JPSJ.54.2037} \BibitemShut {NoStop}%
\bibitem [{\citenamefont {Roati}\ \emph {et~al.}(2008)\citenamefont {Roati},
  \citenamefont {D'Errico}, \citenamefont {Fallani}, \citenamefont {Fattori},
  \citenamefont {Fort}, \citenamefont {Zaccanti}, \citenamefont {Modugno},
  \citenamefont {Modugno},\ and\ \citenamefont {Inguscio}}]{Roati2008}%
  \BibitemOpen
  \bibfield  {author} {\bibinfo {author} {\bibfnamefont {G.}~\bibnamefont
  {Roati}}, \bibinfo {author} {\bibfnamefont {C.}~\bibnamefont {D'Errico}},
  \bibinfo {author} {\bibfnamefont {L.}~\bibnamefont {Fallani}}, \bibinfo
  {author} {\bibfnamefont {M.}~\bibnamefont {Fattori}}, \bibinfo {author}
  {\bibfnamefont {C.}~\bibnamefont {Fort}}, \bibinfo {author} {\bibfnamefont
  {M.}~\bibnamefont {Zaccanti}}, \bibinfo {author} {\bibfnamefont
  {G.}~\bibnamefont {Modugno}}, \bibinfo {author} {\bibfnamefont
  {M.}~\bibnamefont {Modugno}},\ and\ \bibinfo {author} {\bibfnamefont
  {M.}~\bibnamefont {Inguscio}},\ }\bibfield  {title} {\bibinfo {title}
  {Anderson localization of a non-interacting bose--einstein condensate},\
  }\href {https://doi.org/10.1038/nature07071} {\bibfield  {journal} {\bibinfo
  {journal} {Nature (London)}\ }\textbf {\bibinfo {volume} {453}},\ \bibinfo
  {pages} {895} (\bibinfo {year} {2008})}\BibitemShut {NoStop}%
\bibitem [{\citenamefont {Billy}\ \emph {et~al.}(2008)\citenamefont {Billy},
  \citenamefont {Josse}, \citenamefont {Zuo}, \citenamefont {Bernard},
  \citenamefont {Hambrecht}, \citenamefont {Lugan}, \citenamefont
  {Cl{\'e}ment}, \citenamefont {Sanchez-Palencia}, \citenamefont {Bouyer},\
  and\ \citenamefont {Aspect}}]{Billy2008}%
  \BibitemOpen
  \bibfield  {author} {\bibinfo {author} {\bibfnamefont {J.}~\bibnamefont
  {Billy}}, \bibinfo {author} {\bibfnamefont {V.}~\bibnamefont {Josse}},
  \bibinfo {author} {\bibfnamefont {Z.}~\bibnamefont {Zuo}}, \bibinfo {author}
  {\bibfnamefont {A.}~\bibnamefont {Bernard}}, \bibinfo {author} {\bibfnamefont
  {B.}~\bibnamefont {Hambrecht}}, \bibinfo {author} {\bibfnamefont
  {P.}~\bibnamefont {Lugan}}, \bibinfo {author} {\bibfnamefont
  {D.}~\bibnamefont {Cl{\'e}ment}}, \bibinfo {author} {\bibfnamefont
  {L.}~\bibnamefont {Sanchez-Palencia}}, \bibinfo {author} {\bibfnamefont
  {P.}~\bibnamefont {Bouyer}},\ and\ \bibinfo {author} {\bibfnamefont
  {A.}~\bibnamefont {Aspect}},\ }\bibfield  {title} {\bibinfo {title} {Direct
  observation of anderson localization of matter waves in a controlled
  disorder},\ }\href {https://doi.org/10.1038/nature07000} {\bibfield
  {journal} {\bibinfo  {journal} {Nature (London)}\ }\textbf {\bibinfo {volume}
  {453}},\ \bibinfo {pages} {891} (\bibinfo {year} {2008})}\BibitemShut
  {NoStop}%
\bibitem [{\citenamefont {Gross}\ and\ \citenamefont
  {Bloch}(2017)}]{Gross2017}%
  \BibitemOpen
  \bibfield  {author} {\bibinfo {author} {\bibfnamefont {C.}~\bibnamefont
  {Gross}}\ and\ \bibinfo {author} {\bibfnamefont {I.}~\bibnamefont {Bloch}},\
  }\bibfield  {title} {\bibinfo {title} {Quantum simulations with ultracold
  atoms in optical lattices},\ }\href {https://doi.org/10.1126/science.aal3837}
  {\bibfield  {journal} {\bibinfo  {journal} {Science}\ }\textbf {\bibinfo
  {volume} {357}},\ \bibinfo {pages} {995} (\bibinfo {year}
  {2017})}\BibitemShut {NoStop}%
\bibitem [{\citenamefont {Blatt}\ and\ \citenamefont {Roos}(2012)}]{Blatt2012}%
  \BibitemOpen
  \bibfield  {author} {\bibinfo {author} {\bibfnamefont {R.}~\bibnamefont
  {Blatt}}\ and\ \bibinfo {author} {\bibfnamefont {C.~F.}\ \bibnamefont
  {Roos}},\ }\bibfield  {title} {\bibinfo {title} {Quantum simulations with
  trapped ions},\ }\href {https://doi.org/10.1038/nphys2252} {\bibfield
  {journal} {\bibinfo  {journal} {Nat. Phys.}\ }\textbf {\bibinfo {volume}
  {8}},\ \bibinfo {pages} {277} (\bibinfo {year} {2012})}\BibitemShut {NoStop}%
\bibitem [{\citenamefont {Marcuzzi}\ \emph {et~al.}(2017)\citenamefont
  {Marcuzzi}, \citenamefont {Min\'a\ifmmode~\check{r}\else \v{r}\fi{}},
  \citenamefont {Barredo}, \citenamefont {de~L\'es\'eleuc}, \citenamefont
  {Labuhn}, \citenamefont {Lahaye}, \citenamefont {Browaeys}, \citenamefont
  {Levi},\ and\ \citenamefont {Lesanovsky}}]{Marcuzzi2017}%
  \BibitemOpen
  \bibfield  {author} {\bibinfo {author} {\bibfnamefont {M.}~\bibnamefont
  {Marcuzzi}}, \bibinfo {author} {\bibfnamefont {J.~c.~v.}\ \bibnamefont
  {Min\'a\ifmmode~\check{r}\else \v{r}\fi{}}}, \bibinfo {author} {\bibfnamefont
  {D.}~\bibnamefont {Barredo}}, \bibinfo {author} {\bibfnamefont
  {S.}~\bibnamefont {de~L\'es\'eleuc}}, \bibinfo {author} {\bibfnamefont
  {H.}~\bibnamefont {Labuhn}}, \bibinfo {author} {\bibfnamefont
  {T.}~\bibnamefont {Lahaye}}, \bibinfo {author} {\bibfnamefont
  {A.}~\bibnamefont {Browaeys}}, \bibinfo {author} {\bibfnamefont
  {E.}~\bibnamefont {Levi}},\ and\ \bibinfo {author} {\bibfnamefont
  {I.}~\bibnamefont {Lesanovsky}},\ }\bibfield  {title} {\bibinfo {title}
  {Facilitation dynamics and localization phenomena in rydberg lattice gases
  with position disorder},\ }\href
  {https://doi.org/10.1103/PhysRevLett.118.063606} {\bibfield  {journal}
  {\bibinfo  {journal} {Phys. Rev. Lett.}\ }\textbf {\bibinfo {volume} {118}},\
  \bibinfo {pages} {063606} (\bibinfo {year} {2017})}\BibitemShut {NoStop}%
\bibitem [{\citenamefont {Signoles}\ \emph {et~al.}(2021)\citenamefont
  {Signoles}, \citenamefont {Franz}, \citenamefont {Ferracini~Alves},
  \citenamefont {G\"arttner}, \citenamefont {Whitlock}, \citenamefont
  {Z\"urn},\ and\ \citenamefont {Weidem\"uller}}]{Signoles2021}%
  \BibitemOpen
  \bibfield  {author} {\bibinfo {author} {\bibfnamefont {A.}~\bibnamefont
  {Signoles}}, \bibinfo {author} {\bibfnamefont {T.}~\bibnamefont {Franz}},
  \bibinfo {author} {\bibfnamefont {R.}~\bibnamefont {Ferracini~Alves}},
  \bibinfo {author} {\bibfnamefont {M.}~\bibnamefont {G\"arttner}}, \bibinfo
  {author} {\bibfnamefont {S.}~\bibnamefont {Whitlock}}, \bibinfo {author}
  {\bibfnamefont {G.}~\bibnamefont {Z\"urn}},\ and\ \bibinfo {author}
  {\bibfnamefont {M.}~\bibnamefont {Weidem\"uller}},\ }\bibfield  {title}
  {\bibinfo {title} {Glassy dynamics in a disordered heisenberg quantum spin
  system},\ }\href {https://doi.org/10.1103/PhysRevX.11.011011} {\bibfield
  {journal} {\bibinfo  {journal} {Phys. Rev. X}\ }\textbf {\bibinfo {volume}
  {11}},\ \bibinfo {pages} {011011} (\bibinfo {year} {2021})}\BibitemShut
  {NoStop}%
\bibitem [{\citenamefont {Sauerwein}\ \emph {et~al.}(2023)\citenamefont
  {Sauerwein}, \citenamefont {Orsi}, \citenamefont {Uhrich}, \citenamefont
  {Bandyopadhyay}, \citenamefont {Mattiotti}, \citenamefont {Cantat-Moltrecht},
  \citenamefont {Pupillo}, \citenamefont {Hauke},\ and\ \citenamefont
  {Brantut}}]{Sauerwein2023}%
  \BibitemOpen
  \bibfield  {author} {\bibinfo {author} {\bibfnamefont {N.}~\bibnamefont
  {Sauerwein}}, \bibinfo {author} {\bibfnamefont {F.}~\bibnamefont {Orsi}},
  \bibinfo {author} {\bibfnamefont {P.}~\bibnamefont {Uhrich}}, \bibinfo
  {author} {\bibfnamefont {S.}~\bibnamefont {Bandyopadhyay}}, \bibinfo {author}
  {\bibfnamefont {F.}~\bibnamefont {Mattiotti}}, \bibinfo {author}
  {\bibfnamefont {T.}~\bibnamefont {Cantat-Moltrecht}}, \bibinfo {author}
  {\bibfnamefont {G.}~\bibnamefont {Pupillo}}, \bibinfo {author} {\bibfnamefont
  {P.}~\bibnamefont {Hauke}},\ and\ \bibinfo {author} {\bibfnamefont {J.-P.}\
  \bibnamefont {Brantut}},\ }\bibfield  {title} {\bibinfo {title} {Engineering
  random spin models with atoms in a high-finesse cavity},\ }\href
  {https://doi.org/10.1038/s41567-023-02033-3} {\bibfield  {journal} {\bibinfo
  {journal} {Nat. Phys.}\ }\textbf {\bibinfo {volume} {19}},\ \bibinfo {pages}
  {1128} (\bibinfo {year} {2023})}\BibitemShut {NoStop}%
\bibitem [{\citenamefont {Vestg\aa{}rden}\ \emph {et~al.}(2008)\citenamefont
  {Vestg\aa{}rden}, \citenamefont {Bergli},\ and\ \citenamefont
  {Galperin}}]{Galperin2008}%
  \BibitemOpen
  \bibfield  {author} {\bibinfo {author} {\bibfnamefont {J.~I.}\ \bibnamefont
  {Vestg\aa{}rden}}, \bibinfo {author} {\bibfnamefont {J.}~\bibnamefont
  {Bergli}},\ and\ \bibinfo {author} {\bibfnamefont {Y.~M.}\ \bibnamefont
  {Galperin}},\ }\bibfield  {title} {\bibinfo {title} {Nonlinearly driven
  landau-zener transition in a qubit with telegraph noise},\ }\href
  {https://doi.org/10.1103/PhysRevB.77.014514} {\bibfield  {journal} {\bibinfo
  {journal} {Phys. Rev. B}\ }\textbf {\bibinfo {volume} {77}},\ \bibinfo
  {pages} {014514} (\bibinfo {year} {2008})}\BibitemShut {NoStop}%
\bibitem [{\citenamefont {Paladino}\ \emph {et~al.}(2014)\citenamefont
  {Paladino}, \citenamefont {Galperin}, \citenamefont {Falci},\ and\
  \citenamefont {Altshuler}}]{Paladino2014}%
  \BibitemOpen
  \bibfield  {author} {\bibinfo {author} {\bibfnamefont {E.}~\bibnamefont
  {Paladino}}, \bibinfo {author} {\bibfnamefont {Y.~M.}\ \bibnamefont
  {Galperin}}, \bibinfo {author} {\bibfnamefont {G.}~\bibnamefont {Falci}},\
  and\ \bibinfo {author} {\bibfnamefont {B.~L.}\ \bibnamefont {Altshuler}},\
  }\bibfield  {title} {\bibinfo {title} {1/f noise: Implications for
  solid-state quantum information},\ }\href
  {https://doi.org/10.1103/RevModPhys.86.361} {\bibfield  {journal} {\bibinfo
  {journal} {Rev. Mod. Phys.}\ }\textbf {\bibinfo {volume} {86}},\ \bibinfo
  {pages} {361} (\bibinfo {year} {2014})}\BibitemShut {NoStop}%
\bibitem [{\citenamefont {Mutter}\ and\ \citenamefont
  {Burkard}(2022)}]{Mutter2022}%
  \BibitemOpen
  \bibfield  {author} {\bibinfo {author} {\bibfnamefont {P.~M.}\ \bibnamefont
  {Mutter}}\ and\ \bibinfo {author} {\bibfnamefont {G.}~\bibnamefont
  {Burkard}},\ }\bibfield  {title} {\bibinfo {title} {Fingerprints of qubit
  noise in transient cavity transmission},\ }\href
  {https://doi.org/10.1103/PhysRevLett.128.236801} {\bibfield  {journal}
  {\bibinfo  {journal} {Phys. Rev. Lett.}\ }\textbf {\bibinfo {volume} {128}},\
  \bibinfo {pages} {236801} (\bibinfo {year} {2022})}\BibitemShut {NoStop}%
\end{thebibliography}%
\end{document}